\documentclass[9pt,graphicx,subfigure,axodraw]{article}
\usepackage{amsfonts}
\usepackage{amssymb}
\usepackage[usenames,dvipsnames]{color}
\usepackage{graphicx}
\usepackage{amsmath}
\usepackage{amsfonts}
\usepackage{amssymb}
\usepackage{cleveref}
\usepackage{float}
\restylefloat{table}
\usepackage{chngpage}
\def\rmuu{\gamma^{\mu}}
\def\rmud{\gamma_{\mu}}
\def\PL{{1-\gamma_5\over 2}}
\def\PR{{1+\gamma_5\over 2}}
\def\sinW2{\sin^2\theta_W}
\def\AEM{\alpha_{EM}}
\def\mul{M_{\tilde{u} L}^2}
\def\mur{M_{\tilde{u} R}^2}
\def\mdl{M_{\tilde{d} L}^2}
\def\mdr{M_{\tilde{d} R}^2}
\def\mz2{M_{z}^2}
\def\c2b{\cos 2\beta}
\def\au{A_u}
\def\ad{A_d}
\def\cob{\cot \beta}
\def\v#1{v_#1}
\def\tb{\tan\beta}
\def\epem{$e^+e^-$}
\def\KK{$K^0$-$\overline{K^0}$}
\def\wi{\omega_i}
\def\xj{\chi_j}
\def\Wmu{W_\mu}
\def\Wnu{W_\nu}
\def\m#1{{\tilde m}_#1}
\def\mH{m_H}
\def\mw#1{{\tilde m}_{\omega #1}}
\def\mx#1{{\tilde m}_{\chi^{0}_#1}}
\def\mc#1{{\tilde m}_{\chi^{+}_#1}}
\def\mwi{{\tilde m}_{\omega i}}
\def\mxi{{\tilde m}_{\chi^{0}_i}}
\def\mci{{\tilde m}_{\chi^{+}_i}}

\def\ch{{\tilde\chi^{+}_1}}
\def\c2{{\tilde\chi^{+}_2}}

\def\tt{{\tilde\theta}}

\def\tp{{\tilde\phi}}

\def\mz{M_z}
\def\sw{\sin\theta_W}
\def\cw{\cos\theta_W}
\def\cb{\cos\beta}
\def\sb{\sin\beta}
\def\rwi{r_{\omega i}}
\def\rxj{r_{\chi j}}
\def\rfp{r_f'}
\def\Kik{K_{ik}}
\def\Fq2{F_{2}(q^2)}
\def\f{\({\cal F}\)}
\def\d1{{\f(\tilde c;\tilde s;\tilde W)+ \f(\tilde c;\tilde \mu;\tilde W)}}
\def\tw{\tan\theta_W}
\def\sec2w{sec^2\theta_W}
\begin{document}
%This is dvips(k) 5.86 Cobegin{document}
\baselineskip 18pt
%t
\def\today{\ifcase\month\or
 January\or February\or March\or April\or May\or June\or
 July\or August\or September\or October\or November\or December\fi
 \space\number\day, \number\year}
\def\thebibliography#1{\section*{References\markboth
 {References}{References}}\list
 {[\arabic{enumi}]}{\settowidth\labelwidth{[#1]}
 \leftmargin\labelwidth
 \advance\leftmargin\labelsep
 \usecounter{enumi}}
 \def\newblock{\hskip .11em plus .33em minus .07em}
 \sloppy
 \sfcode`\.=1000\relax}
\let\endthebibliography=\endlist
\def\lsim{\ ^<\llap{$_\sim$}\ }
\def\gsim{\ ^>\llap{$_\sim$}\ }
\def\r2{\sqrt 2}
\def\beq{\begin{equation}}
\def\eeq{\end{equation}}
\def\beqn{\begin{eqnarray}}
\def\eeqn{\end{eqnarray}}
\def\rmuu{\gamma^{\mu}}
\def\rmud{\gamma_{\mu}}
\def\PL{{1-\gamma_5\over 2}}
\def\PR{{1+\gamma_5\over 2}}
\def\sinW2{\sin^2\theta_W}
\def\AEM{\alpha_{EM}}
\def\mul{M_{\tilde{u} L}^2}
\def\mur{M_{\tilde{u} R}^2}
\def\mdl{M_{\tilde{d} L}^2}
\def\mdr{M_{\tilde{d} R}^2}
\def\mz2{M_{z}^2}
\def\c2b{\cos 2\beta}
\def\au{A_u}
\def\ad{A_d}
\def\cob{\cot \beta}
\def\v#1{v_#1}
\def\tb{\tan\beta}
\def\epem{$e^+e^-$}
\def\KK{$K^0$-$\bar{K^0}$}
\def\wi{\omega_i}
\def\xj{\chi_j}
\def\Wmu{W_\mu}
\def\Wnu{W_\nu}
\def\m#1{{\tilde m}_#1}
\def\mH{m_H}
\def\mw#1{{\tilde m}_{\omega #1}}
\def\mx#1{{\tilde m}_{\chi^{0}_#1}}
\def\mc#1{{\tilde m}_{\chi^{+}_#1}}
\def\mwi{{\tilde m}_{\omega i}}
\def\mxi{{\tilde m}_{\chi^{0}_i}}
\def\mci{{\tilde m}_{\chi^{+}_i}}
\def\mz{M_z}
\def\sw{\sin\theta_W}
\def\cw{\cos\theta_W}
\def\cb{\cos\beta}
\def\sb{\sin\beta}
\def\rwi{r_{\omega i}}
\def\rxj{r_{\chi j}}
\def\rfp{r_f'}
\def\Kik{K_{ik}}
\def\Fq2{F_{2}(q^2)}
\def\f{\({\cal F}\)}
\def\d1{{\f(\tilde c;\tilde s;\tilde W)+ \f(\tilde c;\tilde \mu;\tilde W)}}
%%%%%%%%%%%%%%%%%%%%%%%%%%%%%%%%%%
\def\tw{\tan\theta_W}
\def\sec2w{sec^2\theta_W}
%%%%%%%%%%%%%%%%%%%%%%%%%%%%%%%%%%
\def\ch{{\tilde\chi^{+}_1}}
\def\c2{{\tilde\chi^{+}_2}}

\def\tt{{\tilde\theta}}

\def\tp{{\tilde\phi}}

\def\mz{M_z}
\def\sw{\sin\theta_W}
\def\cw{\cos\theta_W}
\def\cb{\cos\beta}
\def\sb{\sin\beta}
\def\rwi{r_{\omega i}}
\def\rxj{r_{\chi j}}
\def\rfp{r_f'}
\def\Kik{K_{ik}}
\def\Fq2{F_{2}(q^2)}
\def\f{\({\cal F}\)}
\def\d1{{\f(\tilde c;\tilde s;\tilde W)+ \f(\tilde c;\tilde \mu;\tilde W)}}

\def\b{${\cal{B}}(\mu\to {e} \gamma)$~}

%%%%%%%%%%%%%%%%%%%%%%%%%%%%%%%%%%
\def\tw{\tan\theta_W}
\def\sec2w{sec^2\theta_W}
\newcommand{\pn}[1]{{\color{red}{#1}}}
%%%%%%%%%%%%%%%%%%%%%%%%%%%%%%%%%%

\begin{titlepage}

\begin{center}
{\large {\bf
The Neutron Electric Dipole Moment and Probe of PeV Scale Physics
}}\\
%\vskip 0.5 true cm
\vspace{2cm}
\renewcommand{\thefootnote}
{\fnsymbol{footnote}}
Amin Aboubrahim$^{b}$,  Tarek Ibrahim$^{a}$\footnote{Email: tibrahim@zewailcity.edu.eg}
  and Pran Nath$^{c}$\footnote{Emal: nath@neu.edu}
\vskip 0.5 true cm
\end{center}

\date{Feb 14, 2015}

\noindent
{$^{a}$University of Science and Technology, Zewail City of Science and Technology,}\\
{ 6th of October City, Giza 12588, Egypt\footnote{Permanent address:  Department of  Physics, Faculty of Science,
University of Alexandria, Alexandria, Egypt}\\
}
{$^{b}$Department of Physics, Faculty of Science, Beirut Arab University,
Beirut 11-5020, Lebanon\footnote{Email: amin.b@bau.edu.lb}} \\
{$^{c}$Department of Physics, Northeastern University,
Boston, MA 02115-5000, USA} \\
\vskip 1.0 true cm

\centerline{\bf Abstract}
The experimental limit on the neutron electric dipole moment   is used as a possible probe of new physics
beyond the standard model. Within MSSM we use the current experimental limit on the neutron EDM and
possible future improvement as a probe of high scale SUSY. Quantitative analyses show that scalar
masses as large as a PeV and larger could be probed in improved experiment far above the scales accessible
at future colliders. We also discuss the neutron EDM  as a probe of new physics models beyond MSSM.
Specifically we consider an MSSM extension with a particle content including a vectorlike multiplet. Such an extension
brings in new sources of charge conjugation and parity (CP) violation beyond those  in MSSM. These CP phases contribute to the EDM of the quarks and to
the neutron EDM.  These contributions are analyzed in this work where we include the supersymmetric loop diagrams
involving the neutralinos, charginos, the gluino, squark and mirror squark  exchange diagrams at the one loop level. We also take into account
the contributions from the $W$, $Z$, quark and mirror quark exchanges arising from the mixings of the vectorlike generation with the three
generations.  It is shown that the experimental limit on the neutron EDM can be used to probe such new physics models.
In the future one expects the neutron EDM to improve an order of magnitude or more allowing one to extend the
probe of high scale SUSY and of  new physics models.  For the MSSM the probe of high scales could go
up to and beyond  PeV scale masses.


\medskip

\end{titlepage}

\section{Introduction \label{sec1}}
CP violation provides a window to new physics
[For the early history of CP violation and for reviews see e.g.,\cite{Golub:1994cg,sm,Ibrahim:2007fb,Hewett:2012ns}].
One of the important manifestations of CP violation are that such violations generate
electric dipole moment (EDM) for elementary particle, i.e., for the quarks and leptons.
As is well known the EDM of elementary particles in the standard model  are  very
small. For example, for the electron the EDM is estimated to be $|d_e|\simeq  10^{-38}$ $e$cm.
The electroweak sector of the Standard Model gives  an EDM for the neutron of size $|d_n|\sim  10^{-32}-10^{-31}$
$e$cm. These sizes are too small to be observed in any foreseeable experiment.
The QCD sector of the standard model also produces a non-vanishing EDM for the neutron which is
of size $d_n\sim {\cal O} (10^{-16 \theta})$
$e$cm and satisfaction of Eq.~(1) requires $\theta$ to be of size
$10^{-10}$ or smaller where $\theta$ is QCD phase which enters the QCD Lagrangian as
$(\theta g_s^2/32\pi^2) G_{\mu\nu} \tilde G^{\mu\nu}$.  We assume the absence of such a term
by a symmetry such as the Peccei-Quinn symmetry.
In supersymmetric models there are a variety of new sources of CP violation and typically these
new sources of CP violation lead to EDM of the elementary particles in excess of the observed
limits. This phenomenon is often referred to as the SUSY EDM problem.  Several solutions to this problem
have been suggested in the past such as small CP phases ~\cite{earlywork},  mass suppression  ~\cite{Nath:1991dn}
and the cancellation mechanism~\cite{cancellation1,cancellation2} where various diagrams contributing to the %
EDMs cancel to bring the predicted EDM below the experimental value (for an alternate possibility see~\cite{Babu:1999xf}).
 The recent data from the LHC indicates the Higgs mass to be $\sim 126$ GeV which requires a large loop correction
 to lift the tree level mass to the desired experimental value. The sizable loop correction points to a high SUSY scale
 and
 specifically large scalar masses. In view of this one could turn the indication of a large SUSY scale as a possible
 resolution of the  EDM  problem of supersymmetric models. In fact it has been suggested recently~\cite{McKeen:2013dma,Moroi:2013sfa,Altmannshofer:2013lfa,Ibrahim:2014tba,Dhuria:2013ida},
 that
 one can go further and utilize  the current and future improved data on the EDM limits to probe mass scales far beyond those that
 may be accessible at colliders.  We also note in passing that a large SUSY scale also helps suppress
 flavor changing neutral currents (FCNC)  in supersymmetric
 models and helps stabilize the proton against rapid decay from baryon and lepton number violating  dimension five operators
 in grand unified theories.\\

In this work we will focus on the neutron electric dipole moment of the light quarks which in turn generate
an EDM of the neutron as a probe of high scale physics.
The current experimental limit on the EDM of the neutron is~\cite{Baker:2006ts}
\begin{equation}
|d_n| < 2.9 \times 10^{-26}   e{\rm cm} ~~~(90\% ~{\rm CL}).
\label{1.1}
\end{equation}
Higher sensitivity is expected from experiments in the future~\cite{Ito:2007xd}.
 In our analysis here we  consider the neutron EDM as a probe of high scalar masses within
 the minimal supersymmetric standard model (MSSM) as well as consider
 the neutron EDM as a  probe of an extension of MSSM with a vectorlike generation
 which brings in new sources of CP violation.
 A vectorlike generation is
 anomaly free. Further,  a variety of grand unified models, string and D brane models contain vectorlike generations
  \cite{vectorlike}.  Vectorlike generations have been considered by several authors since their discovery
  would constitute new physics (see, e.g., \cite{Babu:2008ge,Liu:2009cc,Martin:2009bg,Ibrahim:2011im,Ibrahim:2010hv,Ibrahim:2010va,Ibrahim:2008gg,Ibrahim:2009uv,Ibrahim:2012ds,Ibrahim:2014tba,Ibrahim:2014oia,Aboubrahim:2014hya,Aboubrahim:2013yfa,Aboubrahim:2013gfa}).\\

  Quark dipole moment have been examined in MSSM in previous works and a complete analysis at the  one loop level is given in~\cite{cancellation1}.
    Here we compute the EDM of the neutron within an extended MSSM where  the
particle content contains in addition a vectorlike multiplet.
  The outline of the rest of the paper is as follows: In section 2 we give the relevant formulae for the
 extension of MSSM with a vectorlike generation.  In section 3 we  give the interactions of W and Z vector
 bosons with the quarks and  mirror quarks of the extended model.  Interactions of the gluino with quarks, squarks,
 mirror quarks and mirror squarks are given in  section 4. Interactions of the charginos and neutralinos with
 quarks, squarks, mirror quarks and mirror squarks are given in section 5.  An analysis of the electric dipole
 moment operator involving loop contributions from W and Z exchange, gluino exchange, and chargino and
 neutralino exchanges is given in  section 6. A numerical estimate of the EDM of the neutron arising from
 these loop contributions is given in section 7. Here we also discuss the neutron EDM as a probe of PeV scale
 physics. Conclusions are given in section 8. In  section 9 we give details of how the scalar mass square
 matrices are constructed in the extended model with a vectorlike generation.

\section{Extension of MSSM with a Vector Multiplet  \label{sec2}}
In this section we give details of the extension of MSSM  to include a vectorlike  generation. A vectorlike multiplet consists of an  ordinary fourth generation of
  leptons, quarks and their mirrors. A vectorlike generation is anomaly free and thus its inclusion respects the
  good properties of a gauge theory.
 Vectorlike multiplets arise in a variety of unified models some of which could be low-lying.
They have been used recently in a variety of analyses.
In the analysis below we will assume an extended MSSM with just one  vector multiplet.
 Before proceeding further we
define the notation and give a very brief description of the extended model and  a more detailed
description can be found in the previous works mentioned above. Thus the extended MSSM
contains a vectorlike multiplet. To fix notation the three generations of quarks are denoted by

{
\begin{align}
q_{iL}\equiv
 \left(\begin{matrix} t_{i L}\cr
 ~{b}_{iL}  \end{matrix} \right)  \sim \left(3,2,\frac{1}{6}\right) \ ;  ~~ ~t^c_{iL}\sim \left(3^*,1,-\frac{2}{3}\right)\ ;
 %\nonumber\\
 ~~~ b^c_{i L}\sim \left(3^*,1,\frac{1}{3}\right)\ ;
  ~~~i=1,2,3
\label{2}
\end{align}
}
where the properties  under $SU(3)_C\times SU(2)_L\times U(1)_Y$ are also exhibited.
The last entry in the braces such as $(3,2, 1/6)$ is
  the value of the hypercharge
 $Y$ defined so that $Q=T_3+ Y$.  These leptons have $V-A$ interactions.
We can now add a vectorlike multiplet where we have a fourth family of leptons with $V-A$ interactions
whose transformations can be gotten from Eq.~(2) by letting {$i$ run from 1 to 4.}
A vectorlike quark multiplet also has  mirrors and so we consider these mirror
quarks which have $V+A$ interactions.  {The quantum numbers of the mirrors} are given by

{
\begin{align}
Q^c\equiv
 \left(\begin{matrix} B_{ L}^c \cr
 T_L^c\end{matrix}\right)  \sim \left(3^*,2,-\frac{1}{6}\right)\ ;
~~  T_L \sim  \left(3,1,\frac{2}{3}\right)\ ;  ~~   B_L \sim \left(3^*,1,-\frac{1}{3}\right).
\label{3}
\end{align}
}

Interesting new physics arises when we allow mixings of the vectorlike generation with
the three ordinary generations.  Here we focus on the mixing of the mirrors in the vectorlike
generation with the three generations.
Thus the  superpotential of the model allowing for the mixings
among the three ordinary generations and the vectorlike generation is given by

\begin{align}
W&= -\mu \epsilon_{ij} \hat H_1^i \hat H_2^j+\epsilon_{ij}  [y_{1}  \hat H_1^{i} \hat q_{1L} ^{j}\hat b^c_{1L}
 +y_{1}'  \hat H_2^{j}  \hat q_{1L} ^{i}\hat t^c_{1L}
+y_{2}  \hat H_1^{i} \hat Q^c{^{j}}\hat T_{L}
+y_{2}'  \hat H_2^{j} \hat Q^c{^{i}}\hat B_{L}\nonumber \\
 &+y_{3}  \hat H_1^{i} \hat q_{2L} ^{j}\hat b^c_{2L}
 +y_{3}'  \hat H_2^{j}  \hat q_{2L} ^{i}\hat t^c_{2L}
 +y_{4}  \hat H_1^{i} \hat q_{3L} ^{j}\hat b^c_{3L}
 +y_{4}'  \hat H_2^{j}  \hat q_{3L} ^{i}\hat t^c_{3L}] \nonumber \\
&+ h_{3} \epsilon_{ij}  \hat Q^c{^{i}}\hat q_{1L}^{j}+
h_{3}' \epsilon_{ij}  \hat Q^c{^{i}}\hat q_{2L}^{j}+
h_{3}'' \epsilon_{ij}  \hat Q^c{^{i}}\hat q_{3L}^{j}
+h_4 \hat b_{1L}^c \hat B_{L} +h_5 \hat t_{1L}^c \hat T_{L}\nonumber\\
&+h_4'  \hat b_{2L}^c \hat B_{L} +h_5' \hat t_{2L}^c \hat T_{L}
+h_4''  \hat b_{3L}^c \hat B_{L} +h_5'' \hat t_{3L}^c \hat T_{L}
 \ ,
 \label{5}
\end{align}
where $\mu$ is the complex Higgs mixing parameter so that $\mu= |\mu| e^{i\theta_\mu}$.
The mass terms for the ups, mirror ups,  downs and  mirror downs arise from the term
\beq
{\cal{L}}=-\frac{1}{2}\frac{\partial ^2 W}{\partial{A_i}\partial{A_j}}\psi_ i \psi_ j+\text{h.c.},
\label{6}
\eeq
where $\psi$ and $A$ stand for generic two-component fermion and scalar fields.
After spontaneous breaking of the electroweak symmetry, ($\langle H_1^1 \rangle=v_1/\sqrt{2} $ and $\langle H_2^2\rangle=v_2/\sqrt{2}$),
we have the following set of mass terms written in the four-component spinor notation
so that
\beq
-{\cal L}_m= \bar\xi_R^T (M_u) \xi_L +\bar\eta_R^T(M_{d}) \eta_L +\text{h.c.},
\eeq
where the basis vectors in which the mass matrix is written is given by
\begin{gather}
\bar\xi_R^T= \left(\begin{matrix}\bar t_{ R} & \bar T_R & \bar c_{ R}
&\bar u_{R} \end{matrix}\right),\nonumber\\
\xi_L^T= \left(\begin{matrix} t_{ L} &  T_L &  c_{ L}
& u_{ L} \end{matrix}\right),\nonumber\\
\bar\eta_R^T= \left(\begin{matrix}\bar{b}_R & \bar B_R & \bar{s}_R
&\bar{d}_R \end{matrix}\right),\nonumber\\
\eta_L^T= \left(\begin{matrix} {b_ L} &  B_L &  {s_ L}
& {d_ L} \end{matrix}\right),
\end{gather}
and the mass matrix $M_u$ is given by

\beqn
M_u=
 \left(\begin{matrix} y'_1 v_2/\sqrt{2} & h_5 & 0 & 0 \cr
 -h_3 & y_2 v_1/\sqrt{2} & -h_3' & -h_3'' \cr
0&h_5'&y_3' v_2/\sqrt{2} & 0 \cr
0 & h_5'' & 0 & y_4' v_2/\sqrt{2}\end{matrix} \right)\ .
\label{7aa}
\eeqn
We define the matrix element $(22)$ of the mass matrix as $m_T$ so that,
\beqn
m_T= y_2 v_1/\sqrt 2.
\eeqn
The mass matrix is not hermitian and thus one needs bi-unitary transformations to diagonalize it.
We define the bi-unitary transformation so that

\beq
D^{u \dagger}_R (M_u) D^u_L=\text{diag}(m_{u_1},m_{u_2},m_{u_3}, m_{u_4} ).
\label{7a}
\eeq
Under the bi-unitary transformations the basis vectors transform so that
\beqn
 \left(\begin{matrix} t_{R}\cr
 T_{ R} \cr
c_{R} \cr
u_{R} \end{matrix}\right)=D^{u}_R \left(\begin{matrix} u_{1_R}\cr
 u_{2_R}  \cr
u_{3_R} \cr
u_{4_R}\end{matrix}\right), \  \
\left(\begin{matrix} t_{L}\cr
 T_{ L} \cr
c_{L} \cr
u_{L}\end{matrix} \right)=D^{u}_L \left(\begin{matrix} u_{1_L}\cr
 u_{2_L} \cr
u_{3_L} \cr
u_{4_L}\end{matrix}\right) \ .
\label{8}
\eeqn
%{

A similar analysis goes to the down mass matrix $M_d$ where
\beqn
M_d=
 \left(\begin{matrix} y_1 v_1/\sqrt{2} & h_4 & 0 & 0 \cr
 h_3 & y'_2 v_2/\sqrt{2} & h_3' & h_3'' \cr
0&h_4'&y_3 v_1/\sqrt{2} & 0 \cr
0 & h_4'' & 0 & y_4 v_1/\sqrt{2}\end{matrix} \right)\ .
\label{7bb}
\eeqn
In general $h_3, h_4, h_5, h_3', h_4',h_5',  h_3'', h_4'',h_5''$ can be complex and we define their phases
so that

\beqn
h_k= |h_k| e^{i\chi_k}, ~~h_k'= |h_k'| e^{i\chi_k'}, ~~~h_k''= |h_k''| e^{i\chi_k''}\ ;  k=3,4,5\ .
\label{mix}
\eeqn

We introduce now the mass parameter $m_B$ which defines the mass term in the (22) element of the mass matrix of Eq.~(\ref{7bb}) so that
\beqn
m_B=  y_2' v_2/\sqrt 2.
\eeqn
  Next we  consider  the mixing of the down squarks and the charged mirror sdowns.
The mass squared  matrix of the sdown - mirror sdown comes from three sources:  the F term, the
D term of the potential and the soft {SUSY} breaking terms.
Using the  superpotential of  the mass terms arising from it
after the breaking of  the electroweak symmetry are given by
the Lagrangian
\beq
{\cal L}= {\cal L}_F +{\cal L}_D + {\cal L}_{\rm soft}\ ,
\eeq
where   $ {\cal L}_F$ is deduced from $F_i =\partial W/\partial A_i$, and $- {\cal L}_F=V_F=F_i F^{*}_i$ while the ${\cal L}_D$ is given by
\begin{align}
-{\cal L}_D&=\frac{1}{2} m^2_Z \cos^2\theta_W \cos 2\beta \{\tilde t_{ L} \tilde t^*_{ L} -\tilde b_L \tilde b^*_L
+\tilde c_{ L} \tilde c^*_{ L} -\tilde s_L \tilde s^*_L
+\tilde u_{ L} \tilde u^*_{ L} -\tilde d_L \tilde d^*_L \nonumber \\
&+\tilde B_R \tilde B^*_R -\tilde T_R \tilde T^*_R\}
+\frac{1}{2} m^2_Z \sin^2\theta_W \cos 2\beta \{-\frac{1}{3}\tilde t_{ L} \tilde t^*_{ L}
 +\frac{4}{3}\tilde t_{ R} \tilde t^*_{ R}
-\frac{1}{3}\tilde c_{ L} \tilde c^*_{ L}
 +\frac{4}{3}\tilde c_{ R} \tilde c^*_{ R} \nonumber \\
&-\frac{1}{3}\tilde u_{ L} \tilde u^*_{ L}
 +\frac{4}{3}\tilde u_{ R} \tilde u^*_{ R}
+\frac{1}{3}\tilde T_{ R} \tilde T^*_{ R}
 -\frac{4}{3}\tilde T_{ L} \tilde T^*_{ L}
-\frac{1}{3}\tilde b_{ L} \tilde b^*_{ L}
 -\frac{2}{3}\tilde b_{ R} \tilde b^*_{ R}\nonumber\\
&-\frac{1}{3}\tilde s_{ L} \tilde s^*_{ L}
 -\frac{2}{3}\tilde s_{ R} \tilde s^*_{ R}
-\frac{1}{3}\tilde d_{ L} \tilde d^*_{ L}
 -\frac{2}{3}\tilde d_{ R} \tilde d^*_{ R}
+\frac{1}{3}\tilde B_{ R} \tilde B^*_{ R}
 +\frac{2}{3}\tilde B_{ L} \tilde B^*_{ L}
\}.
\label{12}
\end{align}
For ${\cal L}_{\rm soft}$ we assume the following form
\begin{align}
-{\cal L}_{\text{soft}}&= M^2_{\tilde 1 L} \tilde q^{k*}_{1 L} \tilde q^k_{1 L}
+ M^2_{\tilde 2 L} \tilde q^{k*}_{2 L} \tilde q^k_{2 L}
+ M^2_{\tilde 3 L} \tilde q^{k*}_{3 L} \tilde q^k_{3 L}
+ M^2_{\tilde Q} \tilde Q^{ck*} \tilde Q^{ck}
 + M^2_{\tilde t_1} \tilde t^{c*}_{1 L} \tilde t^c_{1 L} \nonumber \\
& + M^2_{\tilde b_1} \tilde b^{c*}_{1 L} \tilde b^c_{1 L}
+ M^2_{\tilde t_2} \tilde t^{c*}_{2 L} \tilde t^c_{2 L}
+ M^2_{\tilde t_3} \tilde t^{c*}_{3 L} \tilde t^c_{3 L}
+ M^2_{\tilde b_2} \tilde b^{c*}_{2 L} \tilde b^c_{2 L}
+ M^2_{\tilde b_3} \tilde b^{c*}_{3 L} \tilde b^c_{3 L}
+ M^2_{\tilde B} \tilde B^*_L \tilde B_L
 +  M^2_{\tilde T} \tilde T^*_L \tilde T_L \nonumber \\
&+\epsilon_{ij} \{y_1 A_{b} H^i_1 \tilde q^j_{1 L} \tilde b^c_{1L}
-y_1' A_{t} H^i_2 \tilde q^j_{1 L} \tilde t^c_{1L}
+y_3 A_{s} H^i_1 \tilde q^j_{2 L} \tilde b^c_{2L}
-y_3' A_{c} H^i_2 \tilde q^j_{2 L} \tilde t^c_{2L}\nonumber\\
&+y_4 A_{d} H^i_1 \tilde q^j_{3 L} \tilde b^c_{3L}
-y_4' A_{u} H^i_2 \tilde q^j_{3 L} \tilde t^c_{3L}
+y_2 A_{T} H^i_1 \tilde Q^{cj} \tilde T_{L}
-y_2' A_{B} H^i_2 \tilde Q^{cj} \tilde B_{L}
+\text{h.c.}\}\ .
\label{13}
\end{align}
Here $M_{\tilde 1 L},  M_{\tilde T}$, etc are the soft masses and $A_t, A_{b}$, etc are the trilinear couplings.
The trilinear couplings are complex  and we define their phases so that
\begin{gather}
A_b= |A_b| e^{i \alpha_{A_b}} \  ,
 ~~A_{t}=  |A_{t}|
 e^{i\alpha_{A_{t}}} \ ,
  \cdots \ .
\end{gather}
From these terms we construct the scalar mass squared matrices.

{

\section{Interaction with W and Z vector bosons  \label{sec3}}

\begin{figure}
\begin{center}
      \includegraphics[scale=.69]{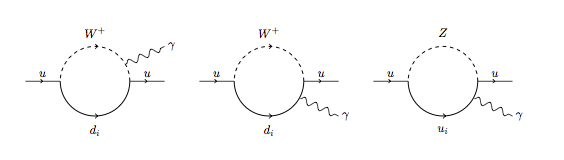}
          \caption{$W$ and $Z$ exchange contributions to the EDM of the up quark.
          Similar exchange contributions exist for the EDM of the down quark where
          $u$ and $d$ are interchanged and $W^+$ is replaced by $W^-$ in the diagrams above.}
\label{feyn1}
\end{center}
\end{figure}

\begin{align}
-{\cal L}_{d W u} &= W^{\dagger}_{\rho}\sum_{i=1}^{4}\sum_{j=1}^{4}\bar{u}_{j}\gamma^{\rho}[G_{L_{ji}}^W P_L + G_{R_{ji}}^W P_R]d_{i}+\text{h.c.},
\end{align}

where
\beqn
G_{L_{ji}}^W= \frac{g}{\sqrt{2}} [D^{u*}_{L4j}D^{d}_{L4i}+
D^{u*}_{L3j}D^{d}_{L3i}+D^{u*}_{L1j}D^{d}_{L1i}],  \\
G_{R_{ji}}^W= \frac{g}{\sqrt{2}}[D^{u*}_{R2j}D^{d}_{R2i}].
\eeqn
For the Z boson exchange the interactions that enter with the up type quarks are given by

\beqn
-{\cal L}_{uu Z} &= Z_{\rho}\sum_{j=1}^{4}\sum_{i=1}^{4}\bar{u}_{j}\gamma^{\rho}[C_{L_{ji}}^{uZ} P_L + C_{R_{ji}}^{uZ} P_R]u_{i},
\eeqn
 where
\beqn
C_{L_{ji}}^{uZ}=\frac{g}{\cos\theta_{W}} [x_1(D^{u*}_{L 4j} D^{u}_{L 4i} +D^{u*}_{L 1j} D^{u}_{L 1i}
+D^{u*}_{L 3j} D^{u}_{L 3i})+ y_1D^{u*}_{L 2j} D^{u}_{L 2i}],
\eeqn
and
\beqn
C_{R_{ji}}^{uZ}=\frac{g}{\cos\theta_{W}} [y_1(D^{u*}_{R 4j} D^{u}_{R 4i} +D^{u*}_{R 1j} D^{u}_{R 1i}
+D^{u*}_{R 3j} D^{u}_{R 3i})+ x_1D^{u*}_{R 2j} D^{u}_{R 2i}],
\eeqn
where
\begin{align}
x_1&=\frac{1}{2} -\frac{2}{3} \sin^2 \theta_W,\\
y_1&=-\frac{2}{3} \sin^2\theta_W.
\end{align}

For the Z boson exchange the interactions that enter with the down type quarks are given by

\beqn
-{\cal L}_{dd Z} &= Z_{\rho}\sum_{j=1}^{4}\sum_{i=1}^{4}\bar{d}_{j}\gamma^{\rho}[C_{L_{ji}}^{dZ} P_L + C_{R_{ji}}^{dZ} P_R]d_{i},
\eeqn
 where
\beqn
C_{L_{ji}}^{dZ}=\frac{g}{\cos\theta_{W}} [x_2(D^{d*}_{L 4j} D^{d}_{L 4i} +D^{d*}_{L 1j} D^{d}_{L 1i}
+D^{d*}_{L 3j} D^{d}_{L 3i})+ y_2D^{d*}_{L 2j} D^{d}_{L 2i}],
\eeqn
and
\beqn
C_{R_{ji}}^{dZ}=\frac{g}{\cos\theta_{W}} [y_2(D^{d*}_{R 4j} D^{d}_{R 4i} +D^{d*}_{R 1j} D^{d}_{R 1i}
+D^{d*}_{R 3j} D^{d}_{R 3i})+ x_2D^{d*}_{R 2j} D^{d}_{R 2i}],
\eeqn
where
\begin{align}
x_2 &=-\frac{1}{2} +\frac{1}{3} \sin^2 \theta_W, \\
y_2 &=\frac{1}{3} \sin^2\theta_W.
\end{align}

\section{Interaction with gluinos  \label{sec4}}

\beqn
-{\cal L}_{qq \tilde{g}} &= \sum_{j=1}^{3}\sum_{k=1}^{3}\sum_{a=1}^{8}\sum_{l=1}^{4}
\sum_{m=1}^{8}\
\bar{q}_{j}[C_{L_{jklm}}^{a} P_L + C_{R_{jklm}}^{a}P_R]
\tilde{g}_a
\tilde{q}^k_{m} + \text{h.c.},
\eeqn
 where
\beqn
C_{L_{jklm}}^{a}=\sqrt{2} g_s T^a_{jk}(D^{q*}_{R2l} \tilde{D}^{q}_{4m}
-D^{q*}_{R4l} \tilde{D}^{q}_{8m}-
D^{q*}_{R3l} \tilde{D}^{q}_{6m}
-D^{q*}_{R1l} \tilde{D}^{q}_{3m}
)e^{-i\xi_3/2},
\eeqn
and
\beqn
C_{R_{jklm}}^{a}=\sqrt{2} g_s T^a_{jk}(D^{q*}_{L4l} \tilde{D}^{q}_{7m}
+D^{q*}_{L3l} \tilde{D}^{q}_{5m}+
D^{q*}_{L1l} \tilde{D}^{q}_{1m}
-D^{q*}_{L2l} \tilde{D}^{q}_{2m}
)e^{i\xi_3/2},
\eeqn
where $\xi_3$ is the phase of the gluino mass.

\section{Interactions with charginos and neutralinos  \label{sec5}}

\begin{figure}
\begin{center}
               \includegraphics[scale=.6]{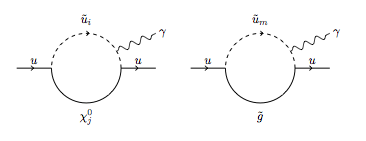}
      \includegraphics[scale=.3]{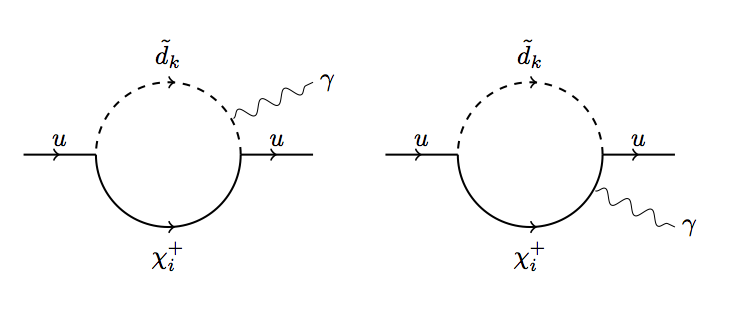}
                \caption{Supersymmetric loop contributions to the EDM of the up-quark. Left panel:
          Loop diagram involving the neutralinos and up-squarks.  Middle left panel: Gluino and
          up-squark loop contribution. Middle right panel: Loop contribution with chargino and d-squark
          exchange with the photon emission from the d-squark line.  Right panel:  Loop contribution with chargino and d-squark
          exchange with the photon emission from the chargino line.           Similar loop contributions exist
          for the EDM of the down quark, where $u$ and $d$ are interchanged, $\tilde u$ and $\tilde d$ are
          interchanged and $\chi^+$ is replaced by $\chi^-$ in the diagrams above.}
\label{figx}
\end{center}
\end{figure}

\noindent
{{
}}\\
 In this section we discuss the  interactions in the mass diagonal basis involving squarks,
 charginos and quarks.  Thus we have
\begin{align}
-{\cal L}_{d-\tilde{u}-\chi^{-}} &= \sum_{j=1}^{4}\sum_{i=1}^{2}\sum_{k=1}^{8}\bar{d}_{j}(C_{jik}^{Ld}P_{L}+C_{jik}^{Rd}P_{R})\tilde{\chi}^{ci}\tilde{u}_{k}+\text{h.c.},
\end{align}
such that,
\begin{align}
%\begin{split}
% \nonumber
C_{jik}^{Ld}=&g(-\kappa_{d}U^{*}_{i2}D^{d*}_{R4j} \tilde{D}^{u}_{7k}
-\kappa_{s}U^{*}_{i2}D^{d*}_{R3j} \tilde{D}^{u}_{5k}-\kappa_{b}U^{*}_{i2}D^{d*}_{R1j} \tilde{D}^{u}_{1k}
-\kappa_{T}U^{*}_{i2}D^{d*}_{R2j} \tilde{D}^{u}_{2k}+U^{*}_{i1}D^{d*}_{R2j} \tilde{D}^{u}_{4k}),\\
%\end{split} \\
%\begin{split}
C_{jik}^{Rd}=&g(-\kappa_{u}V^{}_{i2}D^{d*}_{L4j} \tilde{D}^{u}_{8k}
-\kappa_{c}V^{}_{i2}D^{d*}_{L3j} \tilde{D}^{u}_{6k}
-\kappa_{t}V^{}_{i2}D^{d*}_{L1j} \tilde{D}^{u}_{3k}
-\kappa_{B}V^{}_{i2}D^{d*}_{L2j} \tilde{D}^{u}_{4k}\nonumber\\
&+V^{}_{i1}D^{d*}_{L4j} \tilde{D}^{u}_{7k}
+V^{}_{i1}D^{d*}_{L3j} \tilde{D}^{u}_{5k}
+V^{}_{i1}D^{d*}_{L1j} \tilde{D}^{u}_{1k}),
%\end{split}
\end{align}
and
\begin{align}
-{\cal L}_{u-\tilde{d}-\chi^{-}} &= \sum_{j=1}^{4}\sum_{i=1}^{2}\sum_{k=1}^{8}\bar{u}_{j}(C_{jik}^{Lu}P_{L}+C_{jik}^{Ru}P_{R})\tilde{\chi}^{ci}\tilde{d}_{k}+\text{h.c.},
\end{align}
such that,
\begin{align}
%\begin{split}
% \nonumber
C_{jik}^{Lu}=&g(-\kappa_{u}V^{*}_{i2}D^{u*}_{R4j} \tilde{D}^{d}_{7k}
-\kappa_{c}V^{*}_{i2}D^{u*}_{R3j} \tilde{D}^{d}_{5k}-\kappa_{t}V^{*}_{i2}D^{u*}_{R1j} \tilde{D}^{d}_{1k}
-\kappa_{B}V^{*}_{i2}D^{u*}_{R2j} \tilde{D}^{d}_{2k}+V^{*}_{i1}D^{u*}_{R2j} \tilde{D}^{d}_{4k}),\\
%\end{split} \\
%\begin{split}
C_{jik}^{Ru}=&g(-\kappa_{d}U^{}_{i2}D^{u*}_{L4j} \tilde{D}^{d}_{8k}
-\kappa_{s}U^{}_{i2}D^{u*}_{L3j} \tilde{D}^{d}_{6k}
-\kappa_{b}U^{}_{i2}D^{u*}_{L1j} \tilde{D}^{d}_{3k}
-\kappa_{T}U^{}_{i2}D^{u*}_{L2j} \tilde{D}^{d}_{4k}\nonumber\\
&+U^{}_{i1}D^{u*}_{L4j} \tilde{D}^{d}_{7k}
+U^{}_{i1}D^{u*}_{L3j} \tilde{D}^{d}_{5k}
+U^{}_{i1}D^{u*}_{L1j} \tilde{D}^{d}_{1k}),
%\end{split}
\end{align}
with
\begin{align}
(\kappa_{T},\kappa_{b},\kappa_{s},\kappa_{d})&=\frac{(m_{T},m_{b},m_{s},m_{d})}{\sqrt{2}m_{W}\cos\beta} , \\~\nonumber\\
%\nonumber\\
(\kappa_{B},\kappa_{t},\kappa_{c},\kappa_{u})&=\frac{(m_{B},m_{t},m_{c},m_{u})}{\sqrt{2}m_{W}\sin\beta} .
\end{align}
and
\begin{equation}
U^* M_C V= {\rm diag} (m_{\tilde \chi_1^-}, m_{\tilde \chi_2^-}).
\label{2.4}
\end{equation}
   We now  discuss the  interactions in the mass diagonal basis involving up quarks,
 up squarks and neutralinos.  Thus we have,

\begin{align}
-{\cal L}_{u-\tilde{u}-\chi^{0}} &= \sum_{i=1}^{4}\sum_{j=1}^{4}\sum_{k=1}^{8}\bar{u}_{i}(C_{uijk}^{'L}P_{L}+C_{uijk}^{'R}P_{R})\tilde{\chi}^{0}_{j}\tilde{u}_{k}+\text{h.c.},
\end{align}
such that
\begin{align}
%\begin{split}
C_{uijk}^{'L}=&\sqrt{2}(\alpha_{uj}D^{u *}_{R4i}\tilde{D}^{u}_{7k}-\gamma_{uj}D^{u *}_{R4i}\tilde{D}^{u}_{8k}
+\alpha_{cj}D^{u *}_{R3i}\tilde{D}^{u}_{5k}-\gamma_{cj}D^{u *}_{R3i}\tilde{D}^{u}_{6k}
+\alpha_{tj}D^{u *}_{R1i}\tilde{D}^{u}_{1k}\nonumber\\
&-\gamma_{tj}D^{u *}_{R1i}\tilde{D}^{u}_{3k}
+\beta_{Tj}D^{u *}_{R2i}\tilde{D}^{u}_{4k}-\delta_{Tj}D^{u *}_{R2i}\tilde{D}^{u}_{2k}),
\end{align}
%\end{split} \\  ~\nonumber
%\begin{split}
\begin{align}
C_{uijk}^{'R}=&\sqrt{2}(\beta_{uj}D^{u *}_{L4i}\tilde{D}^{u}_{7k}-\delta_{uj}D^{u *}_{L4i}\tilde{D}^{u}_{8k}
+\beta_{cj}D^{u *}_{L3i}\tilde{D}^{u}_{5k}-\delta_{cj}D^{u *}_{L3i}\tilde{D}^{u}_{6k}
+\beta_{tj}D^{u *}_{L1i}\tilde{D}^{u}_{1k}\nonumber\\
&-\delta_{tj}D^{u *}_{L1i}\tilde{D}^{u}_{3k}
+\alpha_{Tj}D^{u *}_{L2i}\tilde{D}^{u}_{4k}-\gamma_{Tj}D^{u *}_{L2i}\tilde{D}^{u}_{2k})\,,
\end{align}
where

\begin{align}
\alpha_{T j}&=\frac{gm_{T}X^{*}_{3j}}{2m_{W}\cos\beta} \ ;  && \beta_{Tj}=-\frac{2}{3}eX'_{1j}+\frac{g}{\cos\theta_{W}}X'_{2j}\left(-\frac{1}{2}+\frac{2}{3}\sin^{2}\theta_{W}\right) \\
\gamma_{Tj}&=-\frac{2}{3}eX^{'*}_{1j}+\frac{2}{3}\frac{g\sin^{2}\theta_{W}}{\cos\theta_{W}}X^{'*}_{2j} \  ;  && \delta_{Tj}=-\frac{gm_{T}X_{3j}}{2m_{W}\cos\beta}
\end{align}
and

\begin{align}
\alpha_{t j}&=\frac{gm_{t}X_{4j}}{2m_{W}\sin\beta} \ ;  && \alpha_{cj}=\frac{gm_{c}X_{4j}}{2m_{W}\sin\beta} \ ; && \alpha_{uj}=\frac{gm_{u}X_{4j}}{2m_{W}\sin\beta}  \\
\delta_{tj}&=-\frac{gm_{t}X^{*}_{4j}}{2m_{W}\sin\beta} \ ; && \delta_{cj}=-\frac{gm_{c}X^{*}_{4j}}{2m_{W}\sin\beta} \ ; && \delta_{uj}=-\frac{gm_{u}X^{*}_{4j}}{2m_{W}\sin\beta}
\end{align}
and where

\begin{align}
\beta_{tj}=\beta_{cj}=\beta_{uj}&=\frac{2}{3}eX^{'*}_{1j}+\frac{g}{\cos\theta_{W}}X^{'*}_{2j}\left(\frac{1}{2}-\frac{2}{3}\sin^{2}\theta_{W}\right)  \\
\gamma_{tj}=\gamma_{cj}=\gamma_{uj}&=\frac{2}{3}eX'_{1j}-\frac{2}{3}\frac{g\sin^{2}\theta_{W}}{\cos\theta_{W}}X'_{2j}
\end{align}

The interaction of the down quarks, down squarks and neutralinos is given by

\begin{align}
-{\cal L}_{d-\tilde{d}-\chi^{0}} &= \sum_{i=1}^{4}\sum_{j=1}^{4}\sum_{k=1}^{8}\bar{d}_{i}(C_{dijk}^{'L}P_{L}+C_{dijk}^{'R}P_{R})\tilde{\chi}^{0}_{j}\tilde{d}_{k}+\text{h.c.},
\end{align}
such that
\begin{align}
%\begin{split}
C_{dijk}^{'L}=&\sqrt{2}(\alpha_{dj}D^{d *}_{R4i}\tilde{D}^{d}_{7k}-\gamma_{dj}D^{d *}_{R4i}\tilde{D}^{d}_{8k}
+\alpha_{sj}D^{d *}_{R3i}\tilde{D}^{d}_{5k}-\gamma_{sj}D^{d *}_{R3i}\tilde{D}^{d}_{6k}
+\alpha_{bj}D^{d *}_{R1i}\tilde{D}^{d}_{1k}-\gamma_{bj}D^{d *}_{R1i}\tilde{D}^{d}_{3k} \nonumber\\
&+\beta_{Bj}D^{d *}_{R2i}\tilde{D}^{d}_{4k}-\delta_{Bj}D^{d *}_{R2i}\tilde{D}^{d}_{2k}),
\end{align}
%\end{split} \\  ~\nonumber
%\begin{split}
\begin{align}
C_{dijk}^{'R}=&\sqrt{2}(\beta_{dj}D^{d *}_{L4i}\tilde{D}^{d}_{7k}-\delta_{dj}D^{d *}_{L4i}\tilde{D}^{d}_{8k}
+\beta_{sj}D^{d *}_{L3i}\tilde{D}^{d}_{5k}-\delta_{sj}D^{d *}_{L3i}\tilde{D}^{d}_{6k}
+\beta_{bj}D^{d *}_{L1i}\tilde{D}^{d}_{1k}-\delta_{bj}D^{d *}_{L1i}\tilde{D}^{d}_{3k} \nonumber\\
&+\alpha_{Bj}D^{d *}_{L2i}\tilde{D}^{d}_{4k}-\gamma_{Bj}D^{d *}_{L2i}\tilde{D}^{d}_{2k}),
\end{align}
where

\begin{align}
\alpha_{B j}&=\frac{gm_{B}X^{*}_{4j}}{2m_{W}\sin\beta} \ ;  && \beta_{Bj}=\frac{1}{3}eX'_{1j}+\frac{g}{\cos\theta_{W}}X'_{2j}\left(\frac{1}{2}-\frac{1}{3}\sin^{2}\theta_{W}\right) \\
\gamma_{Bj}&=\frac{1}{3}eX^{'*}_{1j}-\frac{1}{3}\frac{g\sin^{2}\theta_{W}}{\cos\theta_{W}}X^{'*}_{2j} \  ;  && \delta_{Bj}=-\frac{gm_{B}X_{4j}}{2m_{W}\sin\beta}
\end{align}
and

\begin{align}
\alpha_{b j}&=\frac{gm_{b}X_{3j}}{2m_{W}\cos\beta} \ ;  && \alpha_{sj}=\frac{gm_{s}X_{3j}}{2m_{W}\cos\beta} \ ; && \alpha_{dj}=\frac{gm_{d}X_{3j}}{2m_{W}\cos\beta}  \\
\delta_{bj}&=-\frac{gm_{b}X^{*}_{3j}}{2m_{W}\cos\beta} \ ; && \delta_{sj}=-\frac{gm_{s}X^{*}_{3j}}{2m_{W}\cos\beta} \ ; && \delta_{dj}=-\frac{gm_{d}X^{*}_{3j}}{2m_{W}\cos\beta}
\end{align}
and where

\begin{align}
\beta_{bj}=\beta_{sj}=\beta_{dj}&=-\frac{1}{3}eX^{'*}_{1j}+\frac{g}{\cos\theta_{W}}X^{'*}_{2j}\left(-\frac{1}{2}+\frac{1}{3}\sin^{2}\theta_{W}\right)  \\
\gamma_{bj}=\gamma_{sj}=\gamma_{dj}&=-\frac{1}{3}eX'_{1j}+\frac{1}{3}\frac{g\sin^{2}\theta_{W}}{\cos\theta_{W}}X'_{2j}
\end{align}
Here $X'$ are defined by

\begin{align}
X'_{1i}&=X_{1i}\cos\theta_{W}+X_{2i}\sin\theta_{W}  \\
X'_{2i}&=-X_ {1i}\sin\theta_{W}+X_{2i}\cos\theta_{W},
\end{align}
where $X$ diagonalizes the neutralino mass matrix and is defined by

 \begin{equation}
X^T M_{\chi^0} X= {\rm diag} \left( m_{\tilde \chi_1^0},  m_{\tilde \chi_2^0}, m_{\tilde \chi_3^0}, m_{\tilde \chi_4^0}\right).
\label{2.8}
\end{equation}

\section{The analysis of Electric Dipole Moment Operator \label{sec6} }

The up  quark will have five different operators arising from the W, Z, gluino, chargino and neutralino contributions as shown in Figs.~\ref{feyn1} and~\ref{figx}. The same thing holds for the down quark. We denote the EDM contributions from these loops by $d_u^W$, $d_u^Z$, $d_u^{\tilde{g}}$, $d_u^{\chi+}$ and $d_u^{\chi0}$, respectively. The same thing holds for the down quarks.

\begin{align}
d_{u}^{W}&=-\frac{1}{16\pi^2}\sum_{i=1}^{4}\frac{m_{d_i}}{m^2_W}\text{Im}(G^{W}_{L 4i}G^{W*}_{R 4i })
\left[I_1\left(\frac{m^{2}_{{d}_{i}}}{m^{2}_{W}}\right) +\frac{1}{3} I_2\left(\frac{m^{2}_{{d}_{i}}}{m^{2}_{W}}\right)
\right],
\end{align}
where the form factor $I_1$  is given by

\begin{align}
I_1(x)&=\frac{2}{(1-x)^{2}}\left[1-\frac{11}{4}x +\frac{1}{4}x^2-\frac{3 x^2\ln x}{2(1-x)} \right].
\end{align}

and
where the form factor $I_2$  is given by

\begin{align}
I_2(x)&=\frac{2}{(1-x)^{2}}\left[1+\frac{1}{4}x +\frac{1}{4}x^2+\frac{3 x\ln x}{2(1-x)} \right].
\label{23}
\end{align}

The W contribution to the down quark EDM is given by
\begin{align}
d_{d}^{W}&=\frac{1}{16\pi^2}\sum_{i=1}^{4}\frac{m_{u_i}}{m^2_W}\text{Im}(G^{W*}_{L i4}G^{W}_{R i4 })
\left[I_1\left(\frac{m^{2}_{{u}_{i}}}{m^{2}_{W}}\right) +\frac{2}{3} I_2\left(\frac{m^{2}_{{u}_{i}}}{m^{2}_{W}}\right)
\right].
\end{align}

The Z exchange contributions to the up and down quarks are given by

\begin{align}
d_{u}^{Z}&=\frac{1}{24\pi^2}\sum_{i=1}^{4}\frac{m_{u_i}}{m^2_Z}\text{Im}(C_{L_{4i}}^{uZ} C_{R_{4i}}^{uZ*})
I_2\left(\frac{m^{2}_{{u}_{i}}}{m^{2}_{Z}}\right)
\end{align}

\begin{align}
d_{d}^{Z}&=-\frac{1}{48\pi^2}\sum_{i=1}^{4}\frac{m_{d_i}}{m^2_Z}\text{Im}(C_{L_{4i}}^{dZ} C_{R_{4i}}^{dZ*})
I_2\left(\frac{m^{2}_{{d}_{i}}}{m^{2}_{Z}}\right)
\end{align}

The gluino contributions to the up and down quarks EDMs are given by

\begin{align}
d_{u}^{\tilde{g}}&=\frac{g^2_s}{9\pi^2}\sum_{m=1}^{8}\frac{m_{\tilde{g}}}{M^2_{\tilde{u}_m}}\text{Im}(K_{L_{um}} K^*_{R_{um}})
B\left(\frac{m^{2}_{\tilde{g}}}{M^{2}_{\tilde{u}_m}}\right)
\end{align}

\begin{align}
d_{d}^{\tilde{g}}&=-\frac{g^2_s}{18\pi^2}\sum_{m=1}^{8}\frac{m_{\tilde{g}}}{M^2_{\tilde{d}_m}}\text{Im}(K_{L_{dm}} K^*_{R_{dm}})
B\left(\frac{m^{2}_{\tilde{g}}}{M^{2}_{\tilde{d}_m}}\right),
\end{align}
where $K_{L_{qm}}$ and $K_{R_{qm}}$ are given by

\beqn
K_{L_{qm}}=(D^{q*}_{R24} \tilde{D}^{q}_{4m}
-D^{q*}_{R44} \tilde{D}^{q}_{8m}-
D^{q*}_{R34} \tilde{D}^{q}_{6m}
-D^{q*}_{R14} \tilde{D}^{q}_{3m}
)e^{-i\xi_3/2}
\eeqn
and
\beqn
K_{R_{qm}}=(D^{q*}_{L44} \tilde{D}^{q}_{7m}
+D^{q*}_{L34} \tilde{D}^{q}_{5m}+
D^{q*}_{L14} \tilde{D}^{q}_{1m}
-D^{q*}_{L24} \tilde{D}^{q}_{2m}
)e^{i\xi_3/2}
\eeqn
and
\begin{equation}
B(x) =  \frac{1}{2(1-x)^2} \left(1+ x + \frac{2 x \ln x}{1-x}\right).
\label{2.6}
\end{equation}

The chargino contribution to the up and down quarks EDMs are given by
\begin{align}
d_{u}^{\chi^{+}}&=\frac{1}{16\pi^2}\sum_{i=1}^{2}\sum_{k=1}^{8}\frac{m_{\chi^{+}_i}}{M^2_{\tilde{d}_{k}}}\text{Im}(C^{Lu}_{4ik}C^{Ru*}_{4ik})
\left[A\left(\frac{m^{2}_{\chi^{+}_i}}{M^{2}_{\tilde{d}_{k}}}\right)
-\frac{1}{3} B\left(\frac{m^{2}_{\chi^{+}_i}}{M^{2}_{\tilde{d}_{k}}}\right)
 \right]
\label{3.1}
\end{align}

\begin{align}
d_{d}^{\chi^{+}}&=\frac{1}{16\pi^2}\sum_{i=1}^{2}\sum_{k=1}^{8}\frac{m_{\chi^{+}_i}}{M^2_{\tilde{u}_{k}}}\text{Im}(C^{Ld}_{4ik}C^{Rd*}_{4ik})
\left[-A\left(\frac{m^{2}_{\chi^{+}_i}}{M^{2}_{\tilde{u}_{k}}}\right)
+\frac{2}{3} B\left(\frac{m^{2}_{\chi^{+}_i}}{M^{2}_{\tilde{u}_{k}}}\right)
 \right],
\label{3.1}
\end{align}

where $A(x)$ is given by
\begin{equation}
A(x)=  \frac{1}{2(1-x)^2} \left(3- x + \frac{2 \ln x}{1-x}\right).
\label{2.2}
\end{equation}

Finally the neutralino contributions are given by

\begin{align}
d_{u}^{\chi^{0}}&=\frac{1}{24\pi^2}\sum_{i=1}^{4}\sum_{k=1}^{8}\frac{m_{\chi^{0}_i}}{M^2_{\tilde{u}_{k}}}\text{Im}(C^{'L}_{u4ik}C^{'R*}_{u4ik})
 B\left(\frac{m^{2}_{\chi^{0}_i}}{M^{2}_{\tilde{u}_{k}}}\right)
\label{3.1}
\end{align}
and

\begin{align}
d_{d}^{\chi^{0}}&=-\frac{1}{48\pi^2}\sum_{i=1}^{4}\sum_{k=1}^{8}\frac{m_{\chi^{0}_i}}{M^2_{\tilde{d}_{k}}}\text{Im}(C^{'L}_{d4ik}C^{'R*}_{d4ik})
 B\left(\frac{m^{2}_{\chi^{0}_{i}}}{M^{2}_{\tilde{d}_{k}}}\right).
\label{3.1}
\end{align}

\section{The neutron EDM  and probe of PeV scale physics\label{sec7}}
To obtain the neutron EDM from the quark EDM, we use the non-relativistic $SU(6)$  quark model which gives
\begin{equation}
d_n= \frac{1}{3} [4 d_d- d_u]\,.
\label{7.1}
\end{equation}
The above value of $d_n$ holds at the electroweak scale and we need to bring it down to the hadronic scale
where it can be compared with experiment. This can be done by evolving $d_n$ from the electroweak scale
down to  the hadronic scale by using renormalization group evolution which gives

\begin{equation}
d_n^E= \eta_E d_n\,,
\label{7.2}
\end{equation}
 where $\eta_E$ is a renomalization group evolution factor.  Numerically it is estimated to be $\sim 1.5$.
We, now, present a numerical analysis of the neutron EDM first for the case of MSSM and next for the
 MSSM extension.  The first analysis involves no mixing with the vectorlike generation and the only
 CP phases that appear are those from the MSSM sector.  Thus in this case all  the mixing parameters, given in Eq.~(\ref{mix}),  are set to zero.
The  second analysis is for the MSSM extension where the mixings of the vectorlike generation with the
three generations are switched on.
In the analysis, in the squark sector we assume $m^{u^2}_0=M^2_{\tilde T}=M^2_{\tilde t_1}=M^2_{\tilde t_2}=M^2_{\tilde t_3}$ and $m^{d^2}_0=M^2_{\tilde 1 L}=M^2_{\tilde B}=M^2_{\tilde b_1}=M^2_{\tilde Q}=M^2_{\tilde 2 L}=M^2_{\tilde b_2}=M^2_{\tilde 3 L}=M^2_{\tilde b_3}$. {{To simplify the numerical analysis further we assume}} $m^u_0=m^d_0=m_0$.  Additionally the trilinear couplings are chosen as such: $A^u_0=A_t=A_T=A_c=A_u$ and $A^d_0=A_b=A_B=A_s=A_d$. \\

As mentioned above first we  explore the possibility of probing high SUSY scales using the neutron EDM
and specifically to see if such scales can lie beyond those that are accessible at colliders.
For this analysis we consider the case when there is no mixing with the vectorlike generation and the neutron EDM
arises from the exchange of the MSSM particles alone which are the charginos, the neutralinos, the gluino and the squarks.
An analysis of this case is presented in fig 3 - fig 5. In fig 3 we display the neutron EDM as a function of the universal scalar mass $m_0$ where the various curves are for values of $\tan\beta$ ranging from
 $5-60$.
The analysis shows that increase in future sensitivities of the neutron EDM will allow us to probe
$m_0$ in the  domain of hundreds of TeV and up to a PeV and even beyond.  This is in contrast to the RUN-II of the
LHC which will allow one to explore the squark masses only in the few TeV region.
Since there are no mixings with the vectorlike generation in this case, the only CP violating phases are
 from the MSSM sector. We discuss the dependence
of the neutron EDM on two of these. In  the left panel of fig 4 we show the dependence of the neutron EDM on the
phase $\xi_3$ of the gluino mass.  We note the very sharp variation of the neutron EDM with $\xi_3$ which
shows that the gluino exchange diagram makes a very significant contribution to the neutron EDM.
The very strong dependence of the neutron EDM
on the gluino exchange diagram is further emphasized in the right panel of
fig 4 where a variation of the neutron EDM
with the  gluino mass is exhibited.  The electroweak sector of the theory also makes a substantial
contribution to the neutron EDM via the chargino, neutralino and the squark exchange diagrams
which involve the phases $\theta_\mu, \alpha_A, \xi_1, \xi_2$ and the electroweak gaugino mass
parameters $m_1, m_2$ and $\mu$.
The dependence of the neutron EDM on $\theta_\mu$ is exhibited in
the left panel of fig 5.
Similar to the left panel of fig 4, this figure too shows a strong dependence of the neutron EDM on the
  CP  phase.  In the right panel of  fig 5 we exhibit  the dependence of $|d_n^E|$ on $\tilde{m}$ where we have
assumed the supergravity boundary condition of the gaugino masses at the electroweak scale, i.e.,
$ m_1 =  \tilde{m},  m_2= 2 \tilde{m},  m_g= 6 \tilde{m}$. Similar to the right panel of
{fig 4 one finds a sharp dependence of $|d_n^E|$ on $\tilde{m}$.\\

 In  table 1 we exhibit the individual contributions of the chargino, neutralino and gluino exchange
 diagrams for two benchmark points. The
 exchange contributions from W and Z vanish because of no mixing with the vectorlike generation
 in this case and are not exhibited.
 The analysis shows that typically the chargino and the gluino contributions are the larger ones
 and the neutralino contribution is suppressed. Further, one finds that typically there is a cancellation
 among the three pieces which reduces the overall size of the quark EDMs. Other choices of the
 phases would lead to other patterns of interference among the terms which explains the rapid phase dependence
 seen, for example, in fig 4 and fig 5.\\

{\tiny

\begin{table}[H]
\begin{center}
\begin{tabular}{l c c c c}
\hline \hline
 & \multicolumn{2}{c}{(i)}& \multicolumn{2}{c}{(ii)}\\ [0.2ex]
\cline{2-3} \cline{4-5}
Contribution & Up & Down & Up & Down\\ [0.5ex]
\hline
Chargino, $d^{\chi^{\pm}}_{q}$ & $3.03\times 10^{-30}$ & $-9.59\times 10^{-28}$ & $4.19\times 10^{-29}$ & $-1.15\times 10^{-26}$ \\ [2ex]
Neutralino, $d^{\chi^{0}}_{q}$ & $-3.69\times 10^{-34}$ & $2.49\times 10^{-31}$ & $-7.53\times 10^{-32}$ & $6.10\times 10^{-29}$\\ [2ex]
Gluino, $d^{g}_{q}$ & $5.14\times 10^{-30}$ & $9.25\times 10^{-29}$ & $7.80\times 10^{-30}$ & $1.40\times 10^{-28}$  \\ [2ex]
Total, $d_{q}$ & $8.16\times 10^{-30}$ & $-8.67\times 10^{-28}$ & $4.96\times 10^{-29}$ & $-1.13\times 10^{-26}$  \\ [2ex]
Total EDM, $|d^{E}_{n}|$ & \multicolumn{2}{c}{$1.77\times 10^{-27}$} &\multicolumn{2}{c}{$2.30\times 10^{-26}$} \\
 [1ex]
\hline
\end{tabular}
\caption{An exhibition of the chargino, neutralino, gluino exchange contributions and their sum for two benchmark points (i) and (ii).
Benchmark (i): $m_g=2$ TeV, $\xi_3=3.3$ and $m_0=m^{u}_{0}=m^{d}_{0}=10$ TeV. Benchmark (ii):  $m_g=30$ TeV, $\xi_3=3.3$ and $m_0=2$ TeV.
The parameter space common between the  two  are: $\tan\beta=25$, $|m_{1}|=70$, $|m_{2}|=200$, $|A^{u}_{0}|=680$, $|A^{d}_{0}|=600$, $|\mu|=350$, $m_{T}=300$, $m_{B}=260$, $|h_{3}|=|h'_{3}|=|h''_{3}|=|h_{4}|=|h'_{4}|=|h''_{4}|=|h_{5}|=|h'_{5}|=|h''_{5}|=0$, $\xi_{1}=2\times 10^{-2}$, $\xi_{2}=2\times 10^{-3}$, $\alpha_{A^{u}_{0}}=2\times 10^{-2}$, $\alpha_{A^{d}_{0}}=3.0$, $\theta_{\mu}=2.6\times 10^{-3}$. All masses, other than $m_0$ and $m_g$, are in GeV, phases in rad and the electric dipole moment in $e$cm.} \label{t1}
\end{center}
\end{table}
}

\begin{figure}[H]
\begin{center}
{\rotatebox{0}{\resizebox*{8cm}{!}{\includegraphics{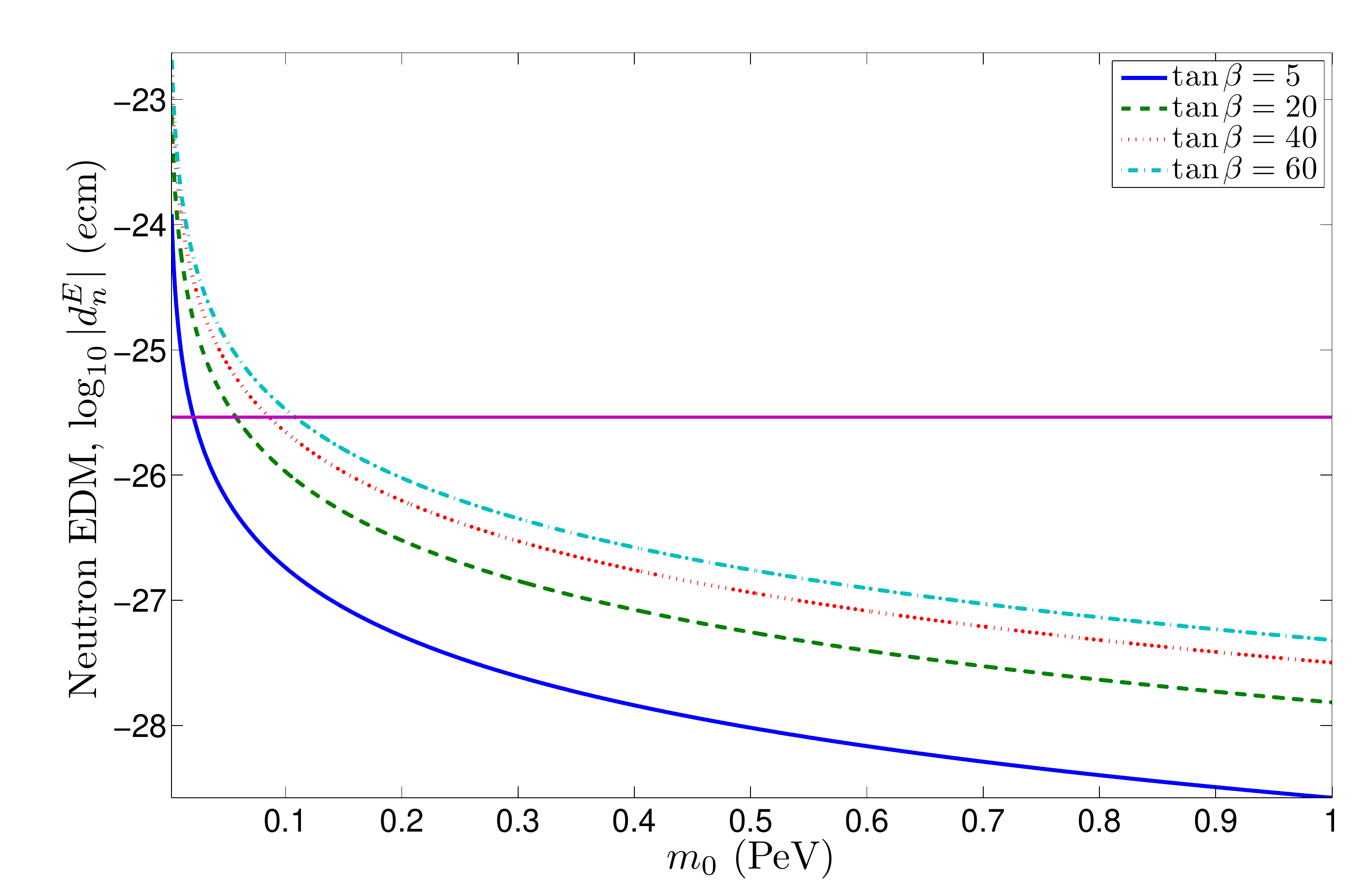}}\hglue5mm}}
\caption{Variation of the neutron EDM $|d^{E}_{n}|$ (log scale) versus $m_0$ ($m_0=m^u_0=m^d_0$) for four values of $\tan\beta$
which are (from bottom to top)  $\tan\beta$ = 5, 20, 40, 60. The common parameters are:  $|m_{1}|=70$, $|m_{2}|=200$, $|A^{u}_{0}|=680$, $|A^{d}_{0}|=600$, $|\mu|=400$, $m_{g}=1000$, $m_{T}=300$, $m_{B}=260$, $|h_{3}|=|h'_{3}|=|h''_{3}|=|h_{4}|=|h'_{4}|=|h''_{4}|=|h_{5}|=|h'_{5}|=|h''_{5}|=0$, $\xi_3=1\times 10^{-3}$, $\xi_1=2\times 10^{-2}$, $\xi_2=2\times 10^{-3}$, $\alpha_{A^{u}_{0}}=2\times 10^{-2}$, $\alpha_{A^{d}_{0}}=3.0$, $\theta_{\mu}=2.0$.  All masses unless otherwise stated are in GeV and phases in rad.
}
\label{fig3}
\end{center}
\end{figure}

\begin{figure}[H]
\begin{center}
{\rotatebox{0}{\resizebox*{7cm}{!}{\includegraphics{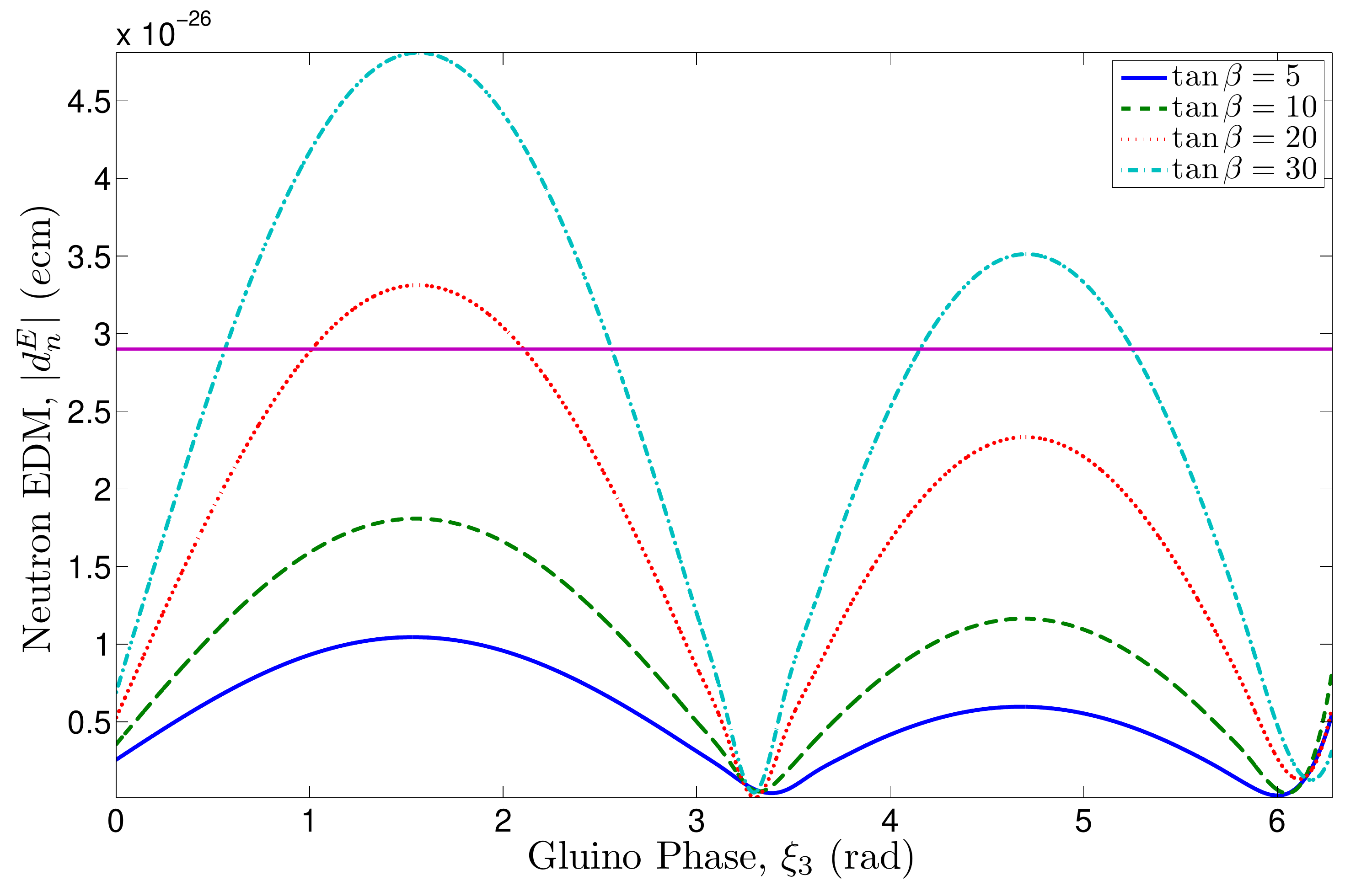}}\hglue5mm}}
{\rotatebox{0}{\resizebox*{7cm}{!}{\includegraphics{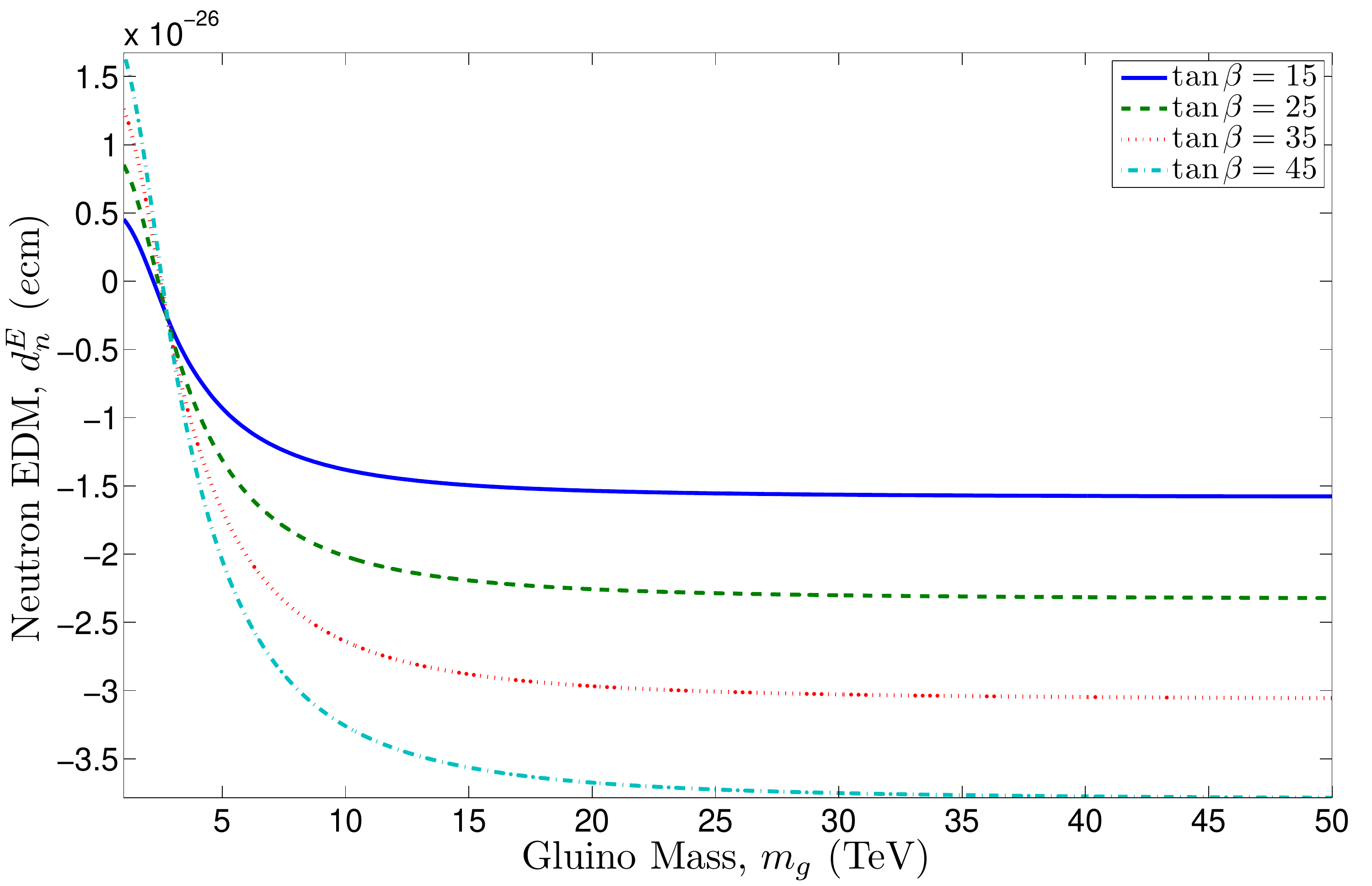}}\hglue5mm}}
\caption{Left panel:
Variation of the neutron EDM $|d^{E}_{n}|$ versus the gluino phase $\xi_3$ for four values of $\tan\beta$. From bottom to top at
 $\xi_3=0$ they are: $\tan\beta$ = 5, 10, 20, 30.  The common parameters are:  $|m_{1}|=70$, $|m_{2}|=200$, $m^{u}_{0}=m^{d}_{0}=3500$, $|A^{u}_{0}|=680$, $|A^{d}_{0}|=600$, $|\mu|=400$, $m_{g}=1000$, $m_{T}=300$, $m_{B}=260$, $|h_{3}|=|h'_{3}|=|h''_{3}|=|h_{4}|=|h'_{4}|=|h''_{4}|=|h_{5}|=|h'_{5}|=|h''_{5}|=0$, $\xi_1=2\times 10^{-2}$, $\xi_2=2\times 10^{-3}$, $\alpha_{A^{u}_{0}}=2\times 10^{-2}$, $\alpha_{A^{d}_{0}}=3.0$, $\theta_{\mu}=2\times 10^{-3}$. All masses are in GeV and phases in rad.
Right panel:
Variation of the neutron EDM $d^{E}_{n}$ versus the gluino mass $m_g$ for four values of $\tan\beta$. From bottom to top at $m_g=50$ TeV
they are: $\tan\beta$ = 45, 35, 25, 15.
The common parameters are:
$|m_{1}|=70$, $|m_{2}|=200$, $m^{u}_{0}=m^{d}_{0}=2000$, $|A^{u}_{0}|=680$, $|A^{d}_{0}|=600$, $|\mu|=350$, $m_{T}=300$, $m_{B}=260$, $|h_{3}|=|h'_{3}|=|h''_{3}|=|h_{4}|=|h'_{4}|=|h''_{4}|=|h_{5}|=|h'_{5}|=|h''_{5}|=0$, $\xi_1=2\times 10^{-2}$, $\xi_2=2\times 10^{-3}$, $\alpha_{A^{u}_{0}}=2\times 10^{-2}$, $\alpha_{A^{d}_{0}}=3.0$, $\theta_{\mu}=2.6\times 10^{-3}$,
$\xi_3=3.3$. All masses, other than $m_g$, are in GeV and phases in rad.}
\label{fig4}
\end{center}
\end{figure}

\begin{figure}[H]
\begin{center}
{\rotatebox{0}{\resizebox*{7.7cm}{!}{\includegraphics{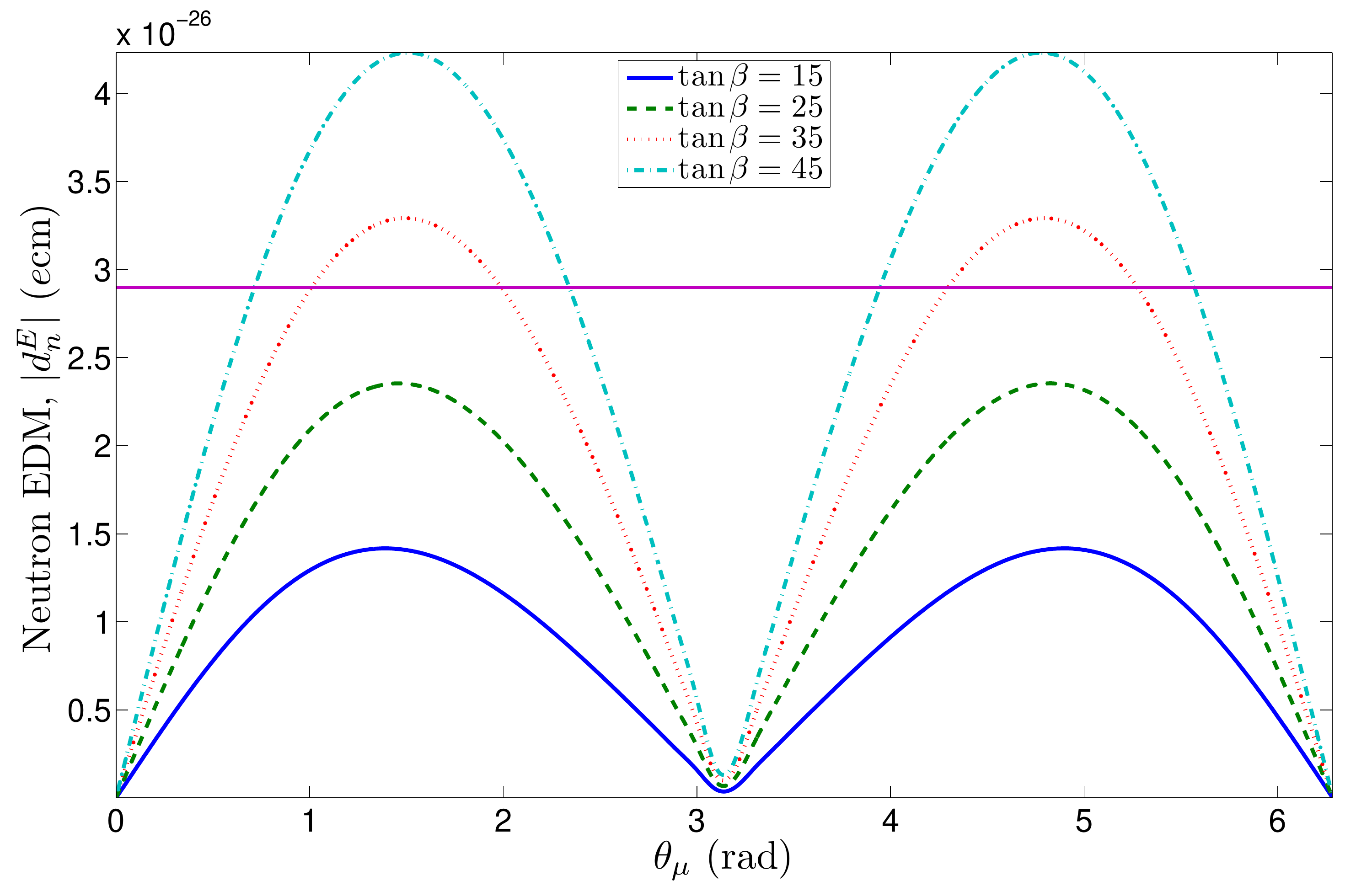}}\hglue5mm}}
{\rotatebox{0}{\resizebox*{6cm}{!}{\includegraphics{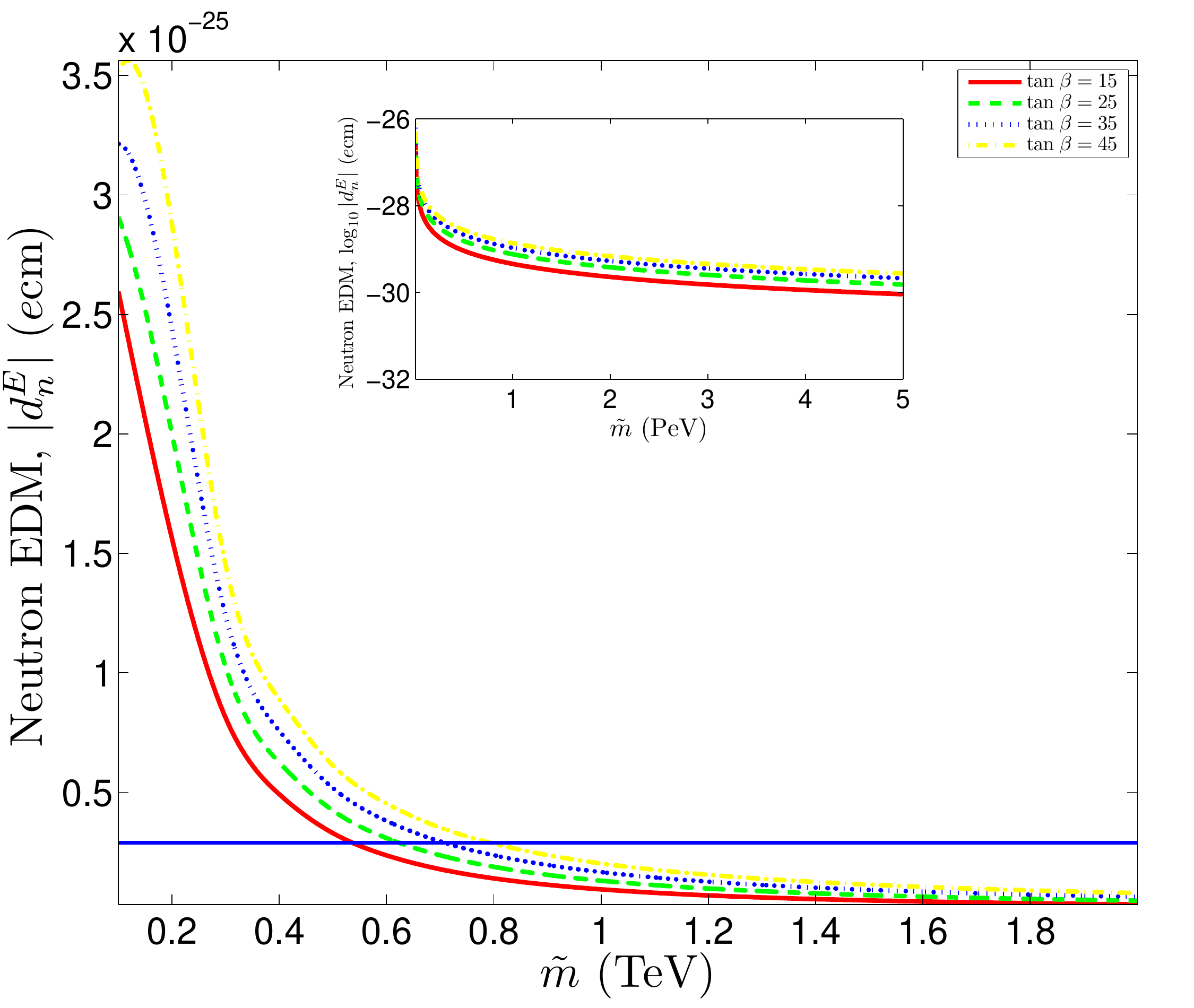}}\hglue5mm}}
\caption{Left panel:
Variation of the neutron EDM $|d^{E}_{n}|$ versus  $\theta_{\mu}$ for four values of $\tan\beta$. From bottom to top they are:
 $\tan\beta$ = 15, 25, 35, 45.  The common  parameters are: $|m_{1}|=70$, $|m_{2}|=200$, $m^{u}_{0}=m^{d}_{0}=80000$, $|A^{u}_{0}|=680$, $|A^{d}_{0}|=600$, $|\mu|=400$, $m_{T}=300$, $m_{B}=260$, $m_g=2000$, $|h_{3}|=|h'_{3}|=|h''_{3}|=|h_{4}|=|h'_{4}|=|h''_{4}|=|h_{5}|=|h'_{5}|=|h''_{5}|=0$, $\xi_1=2\times 10^{-2}$, $\xi_2=2\times 10^{-3}$, $\alpha_{A^{u}_{0}}=2\times 10^{-2}$, $\alpha_{A^{d}_{0}}=2.8$, $\xi_3=3.3$. All masses are in GeV and all phases in rad.
Right panel: Variation of the neutron EDM $|d^{E}_{n}|$ versus $\tilde{m}$  for four values of $\tan\beta$. From bottom to top at
 $\tilde{m}=100$ they are: $\tan\beta$ = 15, 25, 35, 45. The common  parameters are:  $|m_{1}|=\tilde{m}$, $|m_{2}|=2\tilde{m}$, $m_g=6\tilde{m}$, $m^{u}_{0}=m^{d}_{0}=500$, $|A^{u}_{0}|=880$, $|A^{d}_{0}|=600$, $|\mu|=300$, $m_{T}=300$, $m_{B}=260$, $|h_{3}|=|h'_{3}|=|h''_{3}|=|h_{4}|=|h'_{4}|=|h''_{4}|=|h_{5}|=|h'_{5}|=|h''_{5}|=0$, $\xi_1=2\times 10^{-2}$, $\xi_2=2\times 10^{-3}$, $\theta_{\mu}=2\times 10^{-3}$, $\alpha_{A^{u}_{0}}=2\times 10^{-3}$, $\alpha_{A^{d}_{0}}=2.8$, $\xi_3=1\times 10^{-3}$. }
\label{fig5}
\end{center}
\end{figure}

Next we discuss the case when there is mixing between the vector generation and the three generations of
quarks. Here one finds that along with the chargino, neutralino and gluino exchange diagrams, one has
contributions also from the W and Z  exchange diagrams of Fig.~(1). Indeed the contributions
from the W and Z exchange diagrams can be comparable and even larger than the
exchange contributions from the chargino, neutralino and  the gluino.  The relative contributions
from the chargino, the neutralino, the gluino, and from the W and Z bosons are shown in  table 2
for two benchmark points. In this case we note that even for the case when $m_0$ becomes very
large so that the supersymmetric loops give a negligible contribution there will be a non-SUSY
contribution from the exchange of W and Z and of quarks and mirror quarks which will
give a non-vanishing contribution. This is exhibited in fig 6.  Here we note that the EDM
does not fall with increasing  $m_0$ when $m_0$ gets large but rather levels off. The asymptotic
value of the EDM for very large $m_0$, is precisely the contribution from the vectorlike generation.
Obviously the EDM here depends also on the new sources of CP violation such as the phases
$\chi_3, \chi_4, \chi_5''$ in addition to the MSSM phases such as $\theta_\mu$.
 The dependence of $|d_n^E|$ on $\chi_3$ is exhibited in the left panel of   fig 7, on $\chi_4$
 in the right panel of  fig 7,  on $\chi_5''$ in the left panel of fig 8 and on $\theta_\mu$ in the right panel of  fig 8.

\begin{figure}[H]
\begin{center}
{\rotatebox{0}{\resizebox*{7cm}{!}{\includegraphics{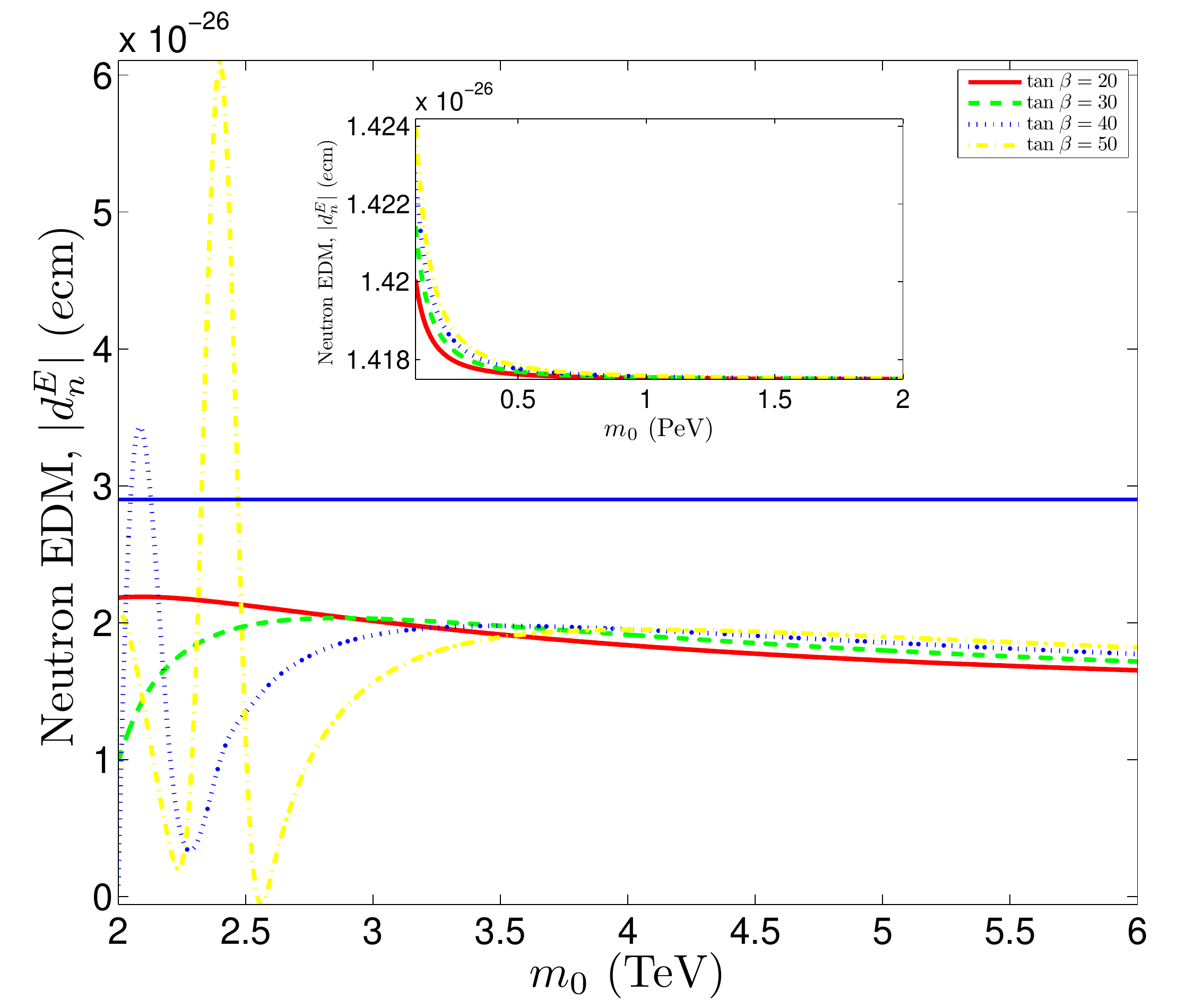}}\hglue5mm}}
\caption{Variation of the neutron EDM $|d^E_{n}|$ versus $m_0$ ($m_0=m^u_0=m^d_0$) for three values of $\tan\beta$.
From bottom to top  at $m_0=6$ TeV, they are: $\tan\beta$ = 20, 30, 40, 50. The common parameters are:  $|m_{1}|=70$, $|m_{2}|=200$, $|A^{u}_{0}|=680$, $|A^{d}_{0}|=600$, $|\mu|=400$, $m_{g}=1000$, $m_{T}=300$, $m_{B}=260$, $|h_{3}|=1.58$, $|h'_{3}|=6.34\times 10^{-2}$, $|h''_{3}|=1.97\times 10^{-2}$, $|h_{4}|=4.42$, $|h'_{4}|=5.07$, $|h''_{4}|=2.87$, $|h_{5}|=6.6$, $|h'_{5}|=2.67$, $|h''_{5}|=1.86\times 10^{-1}$, $\xi_3=1\times 10^{-3}$, $\theta_{\mu}=2.4\times 10^{-3}$, $\xi_1=2\times 10^{-2}$, $\xi_2=2\times 10^{-3}$, $\alpha_{A^{u}_{0}}=2\times 10^{-2}$, $\alpha_{A^{d}_{0}}=3.0$, $\chi_{3}=2\times 10^{-2}$, $\chi'_{3}=1\times 10^{-3}$, $\chi''_{3}=4\times 10^{-3}$, $\chi_{4}=7\times 10^{-3}$, $\chi'_{4}=\chi''_{4}=1\times 10^{-3}$, $\chi_{5}=9\times 10^{-3}$, $\chi'_{5}=5\times 10^{-3}$, $\chi''_{5}=2\times 10^{-3}$.}
\label{fig6}
\end{center}
\end{figure}

\begin{table}[H]
\begin{center}
\begin{tabular}{l c c c c}
\hline \hline
 & \multicolumn{2}{c}{$m_0=3$ TeV}& \multicolumn{2}{c}{$m_0=15$ TeV}\\ [0.2ex]
\cline{2-3} \cline{4-5}
Contribution & Up & Down & Up & Down\\ [0.5ex]
\hline
Chargino, $d^{\chi^{\pm}}_{q}$ & $1.70\times 10^{-28}$ & $-3.01\times 10^{-27}$ & $5.48\times 10^{-30}$ & $-5.14\times 10^{-28}$ \\ [2ex]
Neutralino, $d^{\chi^{0}}_{q}$ & $6.25\times 10^{-31}$ & $2.82\times 10^{-29}$ & $7.37\times 10^{-33}$ & $1.62\times 10^{-31}$ \\ [2ex]
Gluino, $d^{g}_{q}$ & $3.46\times 10^{-29}$ & $-3.93\times 10^{-28}$ & $9.18\times 10^{-32}$ & $-1.05\times 10^{-30}$ \\ [2ex]
W Boson, $d^{W}_{q}$ & $-3.30\times 10^{-28}$ & $-6.49\times 10^{-27}$ & $-3.30\times 10^{-28}$ & $-6.49\times 10^{-27}$ \\ [2ex]
Z Boson, $d^{Z}_{q}$ & $-6.07\times 10^{-29}$ & $-5.58\times 10^{-28}$ & $-6.07\times 10^{-29}$ & $-5.58\times 10^{-28}$ \\ [2ex]
Total, $d_{q}$ & $-1.85\times 10^{-28}$ & $-1.04\times 10^{-26}$ & $-3.85\times 10^{-28}$ & $-7.56\times 10^{-27}$ \\ [2ex]
Total EDM, $|d^{E}_{n}|$ & \multicolumn{2}{c}{$2.12\times 10^{-26}$} &\multicolumn{2}{c}{$1.52\times 10^{-26}$}  \\
 [1ex]
\hline
\end{tabular}
\caption{An exhibition of the chargino, neutralino, gluino, $W$ and $Z$ exchange  contributions to the quark and the neutron EDM and their sum
 for the case when there is mixing of the vectorlike generation with the three generations.  The analysis is for two benchmark
 points with $m_0=3$ TeV and $m_0=15$ TeV.
 The common parameter are: $\tan\beta=40$, $|m_{1}|=70$, $|m_{2}|=200$, $|A^{u}_{0}|=680$, $|A^{d}_{0}|=600$, $|\mu|=400$, $m_{g}=1000$, $m_{T}=300$, $m_{B}=260$, $|h_{3}|=1.58$, $|h'_{3}|=6.34\times 10^{-2}$, $|h''_{3}|=1.97\times 10^{-2}$, $|h_{4}|=4.42$, $|h'_{4}|=5.07$, $|h''_{4}|=2.87$, $|h_{5}|=6.6$, $|h'_{5}|=2.67$, $|h''_{5}|=1.86\times 10^{-1}$, $\xi_3=1\times 10^{-3}$, $\theta_{\mu}=2.6\times 10^{-3}$, $\xi_1=2\times 10^{-2}$, $\xi_2=2\times 10^{-3}$, $\alpha_{A^{u}_{0}}=2\times 10^{-2}$, $\alpha_{A^{d}_{0}}=3.0$, $\chi_{3}=2\times 10^{-2}$, $\chi'_{3}=1\times 10^{-3}$, $\chi''_{3}=4\times 10^{-3}$, $\chi_{4}=7\times 10^{-3}$, $\chi'_{4}=\chi''_{4}=1\times 10^{-3}$, $\chi_{5}=9\times 10^{-3}$, $\chi'_{5}=5\times 10^{-3}$, $\chi''_{5}=2\times 10^{-3}$. All masses are in GeV, all phases in rad and the electric dipole moment in $e$cm.} \label{t2}
\end{center}
\end{table}

\begin{figure}[H]
\begin{center}
{\rotatebox{0}{\resizebox*{6.5cm}{!}{\includegraphics{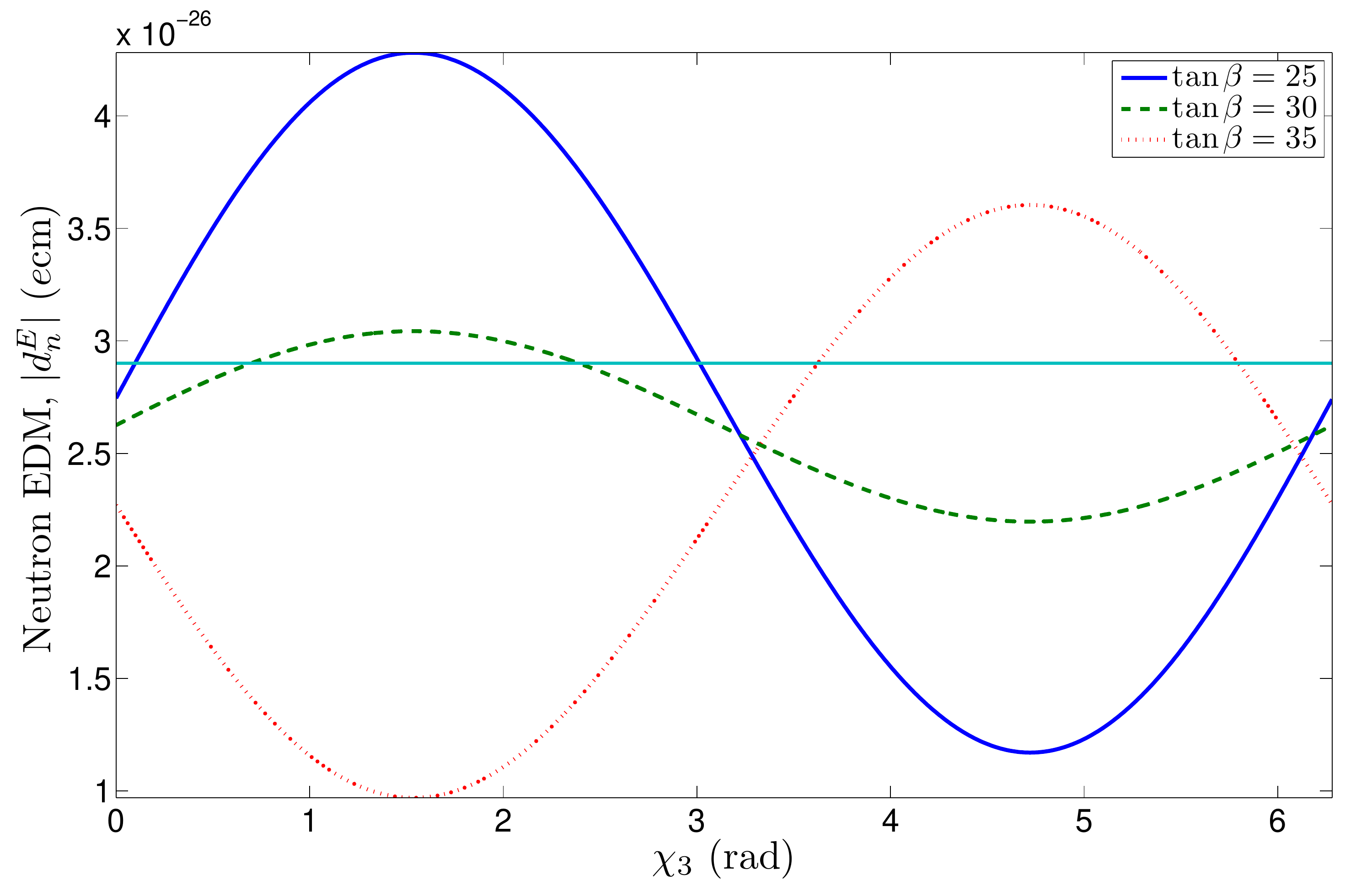}}\hglue5mm}}
{\rotatebox{0}{\resizebox*{6.5cm}{!}{\includegraphics{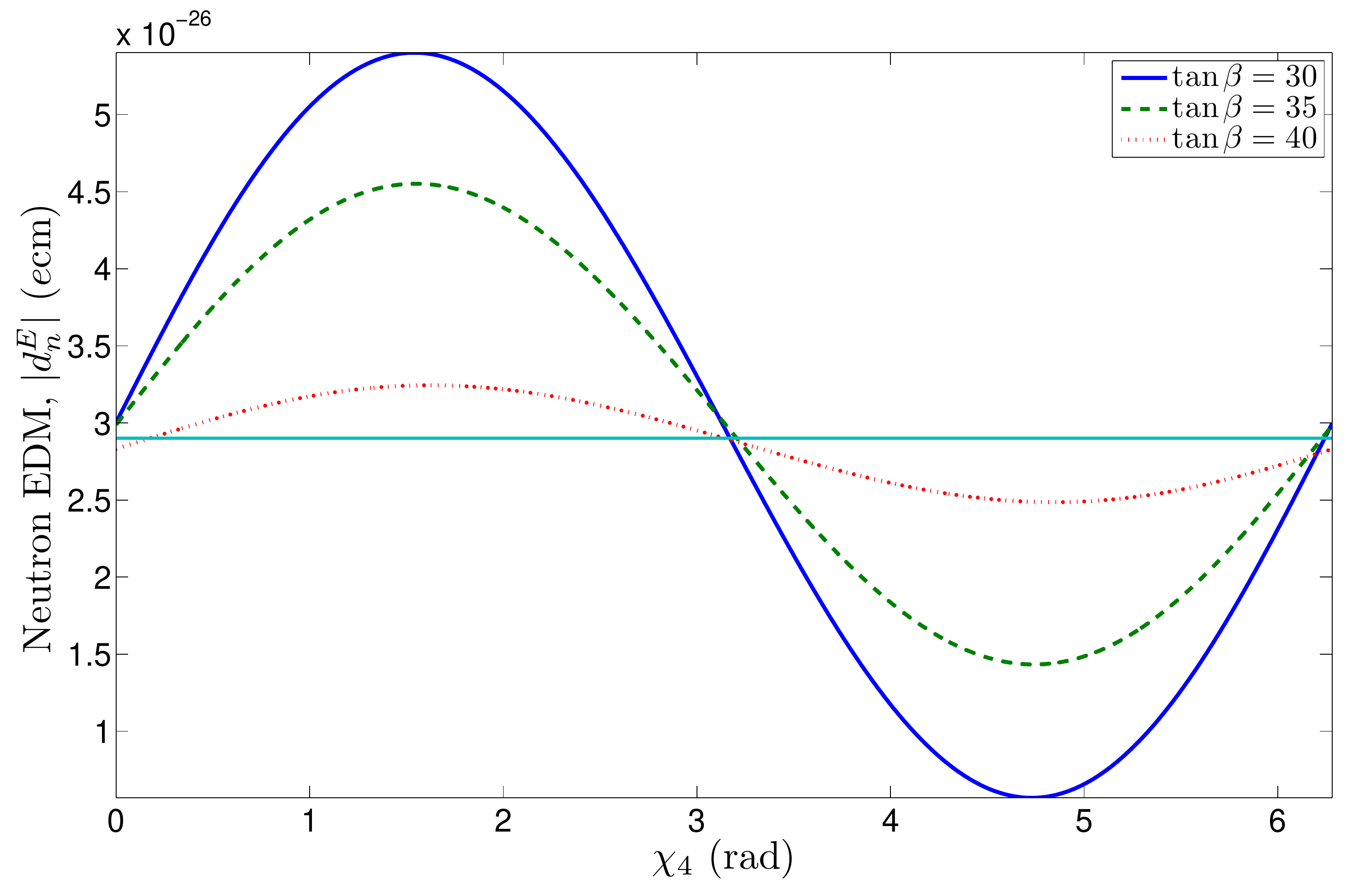}}\hglue5mm}}
\caption{Left panel:
Variation of the neutron EDM $|d^{E}_{n}|$ versus $\chi_{3}$ for three values of $\tan\beta$.
From top to bottom at  $\chi_3=0$ they are:  $\tan\beta$ = 25, 30, 35. The common parameters are: $|m_{1}|=70$, $|m_{2}|=200$, $|\mu|=375$, $|A^{u}_{0}|=680$, $|A^{d}_{0}|=600$, $m^{u}_{0}=m^{d}_{0}=2000$, $m_{g}=1000$, $m_{T}=250$, $m_{B}=240$, $|h_{3}|=1.58$, $|h'_{3}|=6.34\times 10^{-2}$, $|h''_{3}|=1.97\times 10^{-2}$, $|h_{4}|=4.42$, $|h'_{4}|=5.07$, $|h''_{4}|=2.87$, $|h_{5}|=6.6$, $|h'_{5}|=2.67$, $|h''_{5}|=1.86\times 10^{-1}$, $\xi_3=1\times 10^{-3}$, $\xi_1=2\times 10^{-2}$, $\xi_2=2\times 10^{-3}$, $\alpha_{A^{u}_{0}}=4\times 10^{-3}$, $\alpha_{A^{d}_{0}}=1\times 10^{-2}$, $\theta_{\mu}=2.5\times 10^{-3}$, $\chi'_{3}=1\times 10^{-3}$, $\chi''_{3}=4\times 10^{-3}$, $\chi_{4}=7\times 10^{-3}$, $\chi'_{4}=\chi''_{4}=1\times 10^{-3}$, $\chi_{5}=9\times 10^{-3}$, $\chi'_{5}=5\times 10^{-3}$, $\chi''_{5}=2\times 10^{-3}$.
Right panel: Variation of the neutron EDM $|d^{E}_{n}|$ versus $\chi_{4}$ for three values of $\tan\beta$.
From top to bottom  at $\chi_4=0$ they are: $\tan\beta$ = 30, 35, 40.  The common parameters are:
 $|m_{1}|=70$, $|m_{2}|=200$, $|\mu|=300$, $|A^{u}_{0}|=680$, $|A^{d}_{0}|=600$, $m^{u}_{0}=m^{d}_{0}=2000$, $m_{g}=1000$, $m_{T}=m_{B}=260$, $|h_{3}|=1.58$, $|h'_{3}|=6.34\times 10^{-2}$, $|h''_{3}|=1.97\times 10^{-2}$, $|h_{4}|=4.42$, $|h'_{4}|=5.07$, $|h''_{4}|=2.87$, $|h_{5}|=6.6$, $|h'_{5}|=2.67$, $|h''_{5}|=1.86\times 10^{-1}$, $\xi_3=1\times 10^{-3}$, $\xi_1=2\times 10^{-2}$, $\xi_2=2\times 10^{-3}$, $\alpha_{A^{u}_{0}}=2\times 10^{-2}$, $\alpha_{A^{d}_{0}}=1\times 10^{-2}$, $\theta_{\mu}=2.3\times 10^{-3}$, $\chi_{3}=2\times 10^{-2}$, $\chi'_{3}=1\times 10^{-3}$, $\chi''_{3}=4\times 10^{-3}$, $\chi'_{4}=\chi''_{4}=1\times 10^{-3}$, $\chi_{5}=9\times 10^{-3}$, $\chi'_{5}=5\times 10^{-3}$, $\chi''_{5}=2\times 10^{-3}$.}
\label{fig7}
%}\label{fig6}
\end{center}
\end{figure}

\begin{figure}[t]
\begin{center}
{\rotatebox{0}{\resizebox*{6.5cm}{!}{\includegraphics{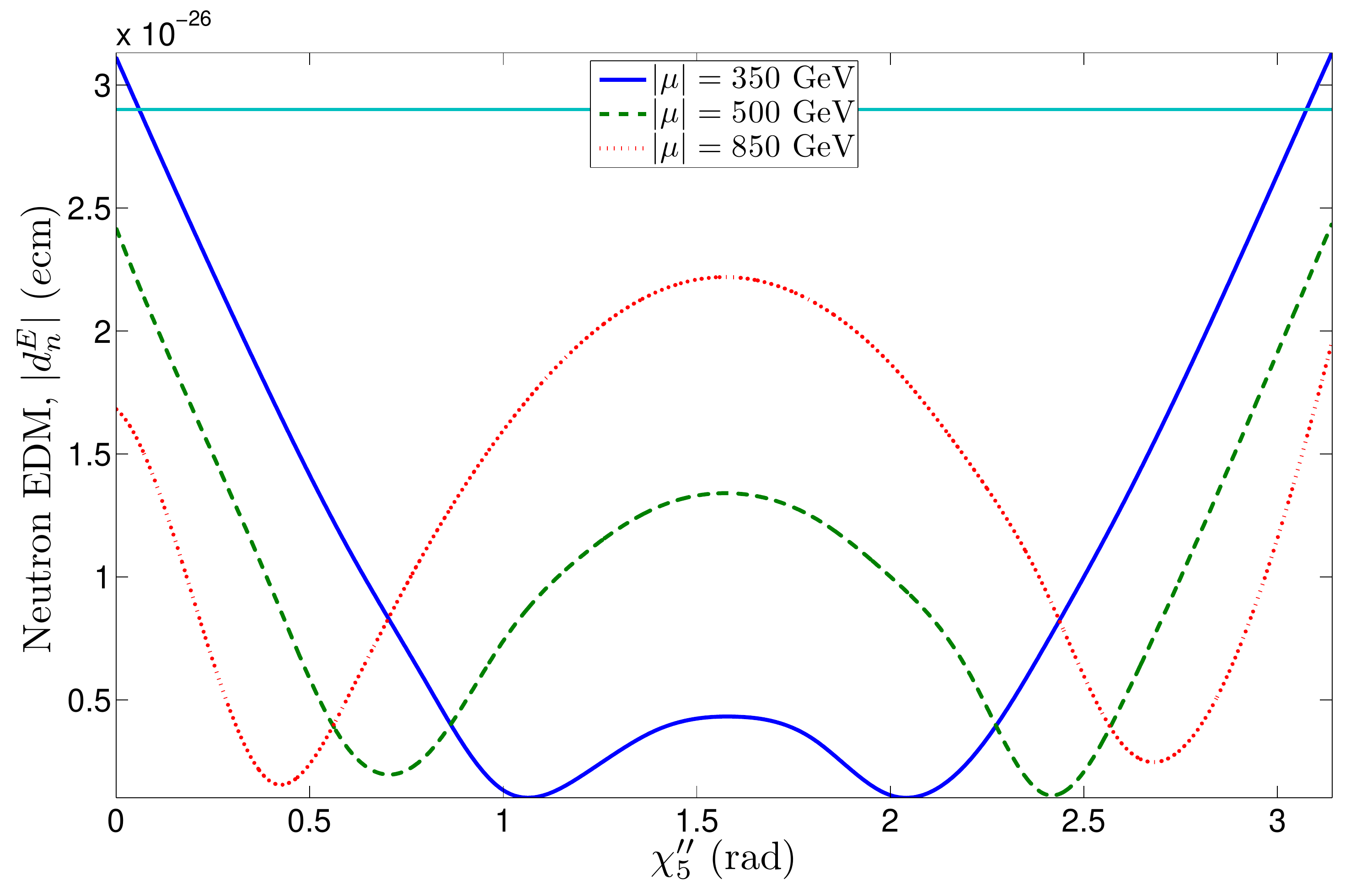}}\hglue5mm}}
{\rotatebox{0}{\resizebox*{6.5cm}{!}{\includegraphics{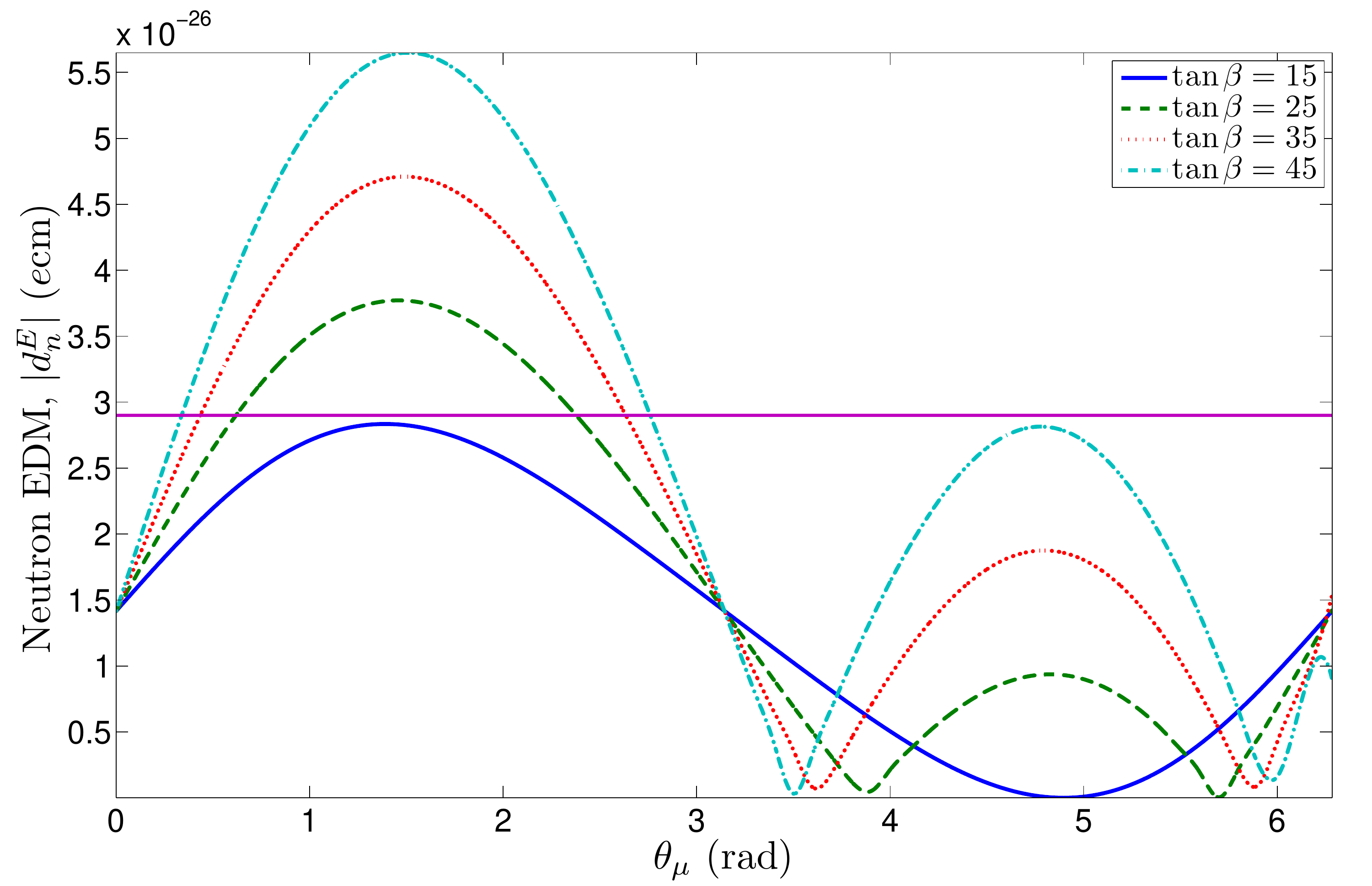}}\hglue5mm}}
\caption{Left panel: Variation of the  neutron EDM $|d^E_{n}|$ versus $\chi''_{5}$ for three values of $|\mu|$.
From bottom to top at $\chi''_5=1.5$ they are: $|\mu|$ = 350, 500, 850. The common parameters are:
 $\tan\beta=15$, $|m_{1}|=70$, $|m_{2}|=200$, $|A^{u}_{0}|=580$, $|A^{d}_{0}|=600$, $m^{u}_{0}=m^{d}_{0}=2000$, $m_{g}=1000$, $m_{T}=250$, $m_{B}=260$, $|h_{3}|=1.58$, $|h'_{3}|=6.34\times 10^{-2}$, $|h''_{3}|=1.97\times 10^{-2}$, $|h_{4}|=4.42$, $|h'_{4}|=5.07$, $|h''_{4}|=2.87$, $|h_{5}|=6.6$, $|h'_{5}|=2.67$, $|h''_{5}|=1.86\times 10^{-1}$, $\xi_3=1\times 10^{-3}$, $\xi_1=2\times 10^{-2}$, $\xi_2=2\times 10^{-3}$, $\alpha_{A^{u}_{0}}=2\times 10^{-2}$, $\alpha_{A^{d}_{0}}=1\times 10^{-2}$, $\theta_{\mu}=3\times 10^{-3}$, $\chi_{3}=2\times 10^{-2}$, $\chi'_{3}=1\times 10^{-3}$, $\chi''_{3}=4\times 10^{-3}$, $\chi_{4}=7\times 10^{-3}$, $\chi'_{4}=\chi''_{4}=1\times 10^{-3}$, $\chi_{5}=9\times 10^{-3}$, $\chi'_{5}=5\times 10^{-3}$.
%}\label{fig9}
Right panel:
Variation of the neutron EDM $|d^E_{n}|$ versus $\theta_{\mu}$ for three values of $\tan\beta$.  From bottom to top at
 $\theta_{\mu}=0$ they are: $\tan\beta$ = 15, 25, 35, 45.  The common parameters are: $|m_{1}|=70$, $|m_{2}|=200$, $m_0=m^u_0=m^d_0=80000$, $|A^{u}_{0}|=680$, $|A^{d}_{0}|=600$, $|\mu|=400$, $m_{g}=2000$, $m_{T}=300$, $m_{B}=260$, $|h_{3}|=1.58$, $|h'_{3}|=6.34\times 10^{-2}$, $|h''_{3}|=1.97\times 10^{-2}$, $|h_{4}|=4.42$, $|h'_{4}|=5.07$, $|h''_{4}|=2.87$, $|h_{5}|=6.6$, $|h'_{5}|=2.67$, $|h''_{5}|=1.86\times 10^{-1}$, $\xi_3=3.3$, $\xi_1=2\times 10^{-2}$, $\xi_2=2\times 10^{-3}$, $\alpha_{A^{u}_{0}}=2\times 10^{-2}$, $\alpha_{A^{d}_{0}}=2.8$, $\chi_{3}=2\times 10^{-2}$, $\chi'_{3}=1\times 10^{-3}$, $\chi''_{3}=4\times 10^{-3}$, $\chi_{4}=7\times 10^{-3}$, $\chi'_{4}=\chi''_{4}=1\times 10^{-3}$, $\chi_{5}=9\times 10^{-3}$, $\chi'_{5}=5\times 10^{-3}$, $\chi''_{5}=2\times 10^{-3}$. }
\label{fig8}
\end{center}
\end{figure}

\section{Conclusion \label{sec8}}
In this work we have investigated the neutron EDM as a possible probe of new physics.  For the case of MSSM
it is shown that the experimental limit on the neutron EDM can be used to probe high scale physics.
Specifically scalar masses as large a PeV and even larger can be probed. We have also investigated
the neutron EDM within an extended MSSM where the particle content of the model contains in addition
a vectorlike multiplet. In section 6 we have given a complete analytic analysis of the neutron EDM which
contains all the relevant diagrams at the one loop level including both the supersymmetric as well as the non-supersymmetric loops. Thus the analysis includes  loops involving exchanges of charginos,
neutralinos, gluino, squarks and mirror squarks. In addition the analysis includes $W$ and $Z$ exchange diagrams
with exchange of quarks and mirror quarks. The vectorlike generation brings in new sources of CP violation which contribute to the quark EDMs. It is shown that in the absence of the cancellation mechanism the experimental limit on the neutron EDM acts as a probe of new physics. Specifically it is shown that assuming CP phases to be ${\cal O}(1)$, and with no cancellation mechanism at work, one can probe scalar masses up to the PeV scale for the MSSM case.
Further, it is shown that the neutron EDM also acts as a probe of the extended MSSM model which
includes a vectorlike multiplet in its particle content.  One expects significant improvements in the sensitivity of
the measurement of the neutron EDM in the future. Thus, for instance, the nEDM Collaboration plans to
measure the neutron EDM with an accuracy of $\sim 9 \times 10^{-28}$ $e$cm which is more than an order of
magnitude better than the current limit. Such an improvement  will allow one to extend the probe of new physics
even further~\cite{Tsentalovich:2014mfa}. For a recent analysis of the neutron EDM in a different class of models see Ref.~\cite{Y.V.Stadnik:2014}. \\

\noindent
{\bf Acknowledgments:}
PN's research  is  supported in part by the NSF grant PHY-1314774.\\

\noindent
\section{Appendix: Mass squared matrices for the scalars  \label{sec9}}

We define the scalar mass squared   matrix $M^2_{\tilde{d}}$  in the basis $(\tilde  b_L, \tilde B_L, \tilde b_R,
\tilde B_R, \tilde s_L, \tilde s_R, \tilde d_L, \tilde d_R)$. We  label the matrix  elements of these as $(M^2_{\tilde d})_{ij}= M^2_{ij}$ where the elements of the matrix are given by
\begin{align}
M^2_{11}&= M^2_{\tilde 1 L}+\frac{v^2_1|y_1|^2}{2} +|h_3|^2 -m^2_Z \cos 2 \beta \left(\frac{1}{2}-\frac{1}{3}\sin^2\theta_W\right), \nonumber\\
M^2_{22}&=M^2_{\tilde B}+\frac{v^2_2|y'_2|^2}{2}+|h_4|^2 +|h'_4|^2+|h''_4|^2 +\frac{1}{3}m^2_Z \cos 2 \beta \sin^2\theta_W, \nonumber\\
M^2_{33}&= M^2_{\tilde b_1}+\frac{v^2_1|y_1|^2}{2} +|h_4|^2 -\frac{1}{3}m^2_Z \cos 2 \beta \sin^2\theta_W, \nonumber\\
M^2_{44}&=  M^2_{\tilde Q}+\frac{v^2_2|y'_2|^2}{2} +|h_3|^2 +|h'_3|^2+|h''_3|^2 +m^2_Z \cos 2 \beta \left(\frac{1}{2}-\frac{1}{3}\sin^2\theta_W\right), \nonumber
\end{align}
\begin{align}
M^2_{55}&=M^2_{\tilde 2 L} +\frac{v^2_1|y_3|^2}{2} +|h'_3|^2 -m^2_Z \cos 2 \beta \left(\frac{1}{2}-\frac{1}{3}\sin^2\theta_W\right), \nonumber\\
M^2_{66}&= M^2_{\tilde b_2}+\frac{v^2_1|y_3|^2}{2}+|h'_4|^2  -\frac{1}{3}m^2_Z \cos 2 \beta \sin^2\theta_W,\nonumber\\
M^2_{77}&=M^2_{\tilde 3 L}+\frac{v^2_1|y_4|^2}{2}+|h''_3|^2-m^2_Z \cos 2 \beta \left(\frac{1}{2}-\frac{1}{3}\sin^2\theta_W\right),  \nonumber\\
M^2_{88}&= M^2_{\tilde b_3}+\frac{v^2_1|y_4|^2}{2}+|h''_4|^2   -\frac{1}{3}m^2_Z \cos 2 \beta \sin^2\theta_W\ . \nonumber
\end{align}

\begin{align}
M^2_{12}&=M^{2*}_{21}=\frac{ v_2 y'_2h^*_3}{\sqrt{2}} +\frac{ v_1 h_4 y^*_1}{\sqrt{2}} ,
M^2_{13}=M^{2*}_{31}=\frac{y^*_1}{\sqrt{2}}(v_1 A^*_{b} -\mu v_2),
M^2_{14}=M^{2*}_{41}=0,\nonumber\\
 M^2_{15} &=M^{2*}_{51}=h'_3 h^*_3,
 M^{2*}_{16}= M^{2*}_{61}=0,  M^{2*}_{17}= M^{2*}_{71}=h''_3 h^*_3,  M^{2*}_{18}= M^{2*}_{81}=0,\nonumber\\
M^2_{23}&=M^{2*}_{32}=0,
M^2_{24}=M^{2*}_{42}=\frac{y'^*_2}{\sqrt{2}}(v_2 A^*_{B} -\mu v_1),  M^2_{25} = M^{2*}_{52}= \frac{ v_2 h'_3y'^*_2}{\sqrt{2}} +\frac{ v_1 y_3 h^*_4}{\sqrt{2}} ,\nonumber\\
 M^2_{26} &=M^{2*}_{62}=0,  M^2_{27} =M^{2*}_{72}=  \frac{ v_2 h''_3y'^*_2}{\sqrt{2}} +\frac{ v_1 y_4 h''^*_4}{\sqrt{2}},  M^2_{28} =M^{2*}_{82}=0, \nonumber\\
M^2_{34}&=M^{2*}_{43}= \frac{ v_2 h_4 y'^*_2}{\sqrt{2}} +\frac{ v_1 y_1 h^*_3}{\sqrt{2}}, M^2_{35} =M^{2*}_{53} =0, M^2_{36} =M^{2*}_{63}=h_4 h'^*_4,\nonumber\\
 M^2_{37} &=M^{2*}_{73} =0,  M^2_{38} =M^{2*}_{83} =h_4 h''^*_4,\nonumber\\
M^2_{45}&=M^{2*}_{54}=0, M^2_{46}=M^{2*}_{64}=\frac{ v_2 y'_2 h'^*_4}{\sqrt{2}} +\frac{ v_1 h'_3 y^*_3}{\sqrt{2}}, \nonumber\\
 M^2_{47} &=M^{2*}_{74}=0,  M^2_{48} =M^{2*}_{84}=  \frac{ v_2 y'_2h''^*_4}{\sqrt{2}} +\frac{ v_1 h''_3 y^*_4}{\sqrt{2}},\nonumber\\
M^2_{56}&=M^{2*}_{65}=\frac{y^*_3}{\sqrt{2}}(v_1 A^*_{s} -\mu v_2),
 M^2_{57} =M^{2*}_{75}=h''_3 h'^*_3,  \nonumber\\
 M^2_{58} &=M^{2*}_{85}=0,  M^2_{67} =M^{2*}_{76}=0,\nonumber\\
 M^2_{68} &=M^{2*}_{86}=h'_4 h''^*_4,  M^2_{78}=M^{2*}_{87}=\frac{y^*_4}{\sqrt{2}}(v_1 A^*_{d} -\mu v_2)\ . \nonumber
\label{14}
\end{align}

We can diagonalize this hermitian mass squared  matrix  by the
 unitary transformation
\begin{gather}
 \tilde D^{d \dagger} M^2_{\tilde d} \tilde D^{d} = diag (M^2_{\tilde d_1},
M^2_{\tilde d_2}, M^2_{\tilde d_3},  M^2_{\tilde d_4},  M^2_{\tilde d_5},  M^2_{\tilde d_6},  M^2_{\tilde d_7},  M^2_{\tilde d_8} )\ .
\end{gather}

%%%%%%%%%%%%%%%%%%%%%%%%%%%%%%%%%%%%%%%%%%%%%%%%%%%%%%%%

 Next we write the   mass$^2$  matrix in the up squark sector the basis $(\tilde  t_{ L}, \tilde T_L,$
$ \tilde t_{ R}, \tilde T_R, \tilde  c_{ L},\tilde c_{ R}, \tilde u_{ L}, \tilde u_{R} )$.
 Thus here we denote the up squark sector mass$^2$ matrix in the form
$(M^2_{\tilde u})_{ij}=m^2_{ij}$ where

\begin{align}
m^2_{11}&= M^2_{\tilde 1 L}+\frac{v^2_2|y'_1|^2}{2} +|h_3|^2 +m^2_Z \cos 2 \beta \left(\frac{1}{2}-\frac{2}{3}\sin^2\theta_W\right), \nonumber\\
m^2_{22}&=M^2_{\tilde T}+\frac{v^2_1|y_2|^2}{2}+|h_5|^2 +|h'_5|^2+|h''_5|^2 -\frac{2}{3}m^2_Z \cos 2 \beta \sin^2\theta_W, \nonumber\\
m^2_{33}&= M^2_{\tilde t_1}+\frac{v^2_2|y'_1|^2}{2} +|h_5|^2 +\frac{2}{3}m^2_Z \cos 2 \beta \sin^2\theta_W, \nonumber\\
m^2_{44}&=  M^2_{\tilde Q}+\frac{v^2_1|y_2|^2}{2} +|h_3|^2 +|h'_3|^2+|h''_3|^2 -m^2_Z \cos 2 \beta \left(\frac{1}{2}-\frac{2}{3}\sin^2\theta_W\right), \nonumber
\end{align}
\begin{align}
m^2_{55}&=M^2_{\tilde 2 L} +\frac{v^2_2|y'_3|^2}{2} +|h'_3|^2 +m^2_Z \cos 2 \beta \left(\frac{1}{2}-\frac{2}{3}\sin^2\theta_W\right), \nonumber\\
m^2_{66}&= M^2_{\tilde t_2}+\frac{v^2_2|y'_3|^2}{2}+|h'_5|^2  +\frac{2}{3}m^2_Z \cos 2 \beta \sin^2\theta_W,\nonumber\\
m^2_{77}&=M^2_{\tilde 3 L}+\frac{v^2_2|y'_4|^2}{2}+|h''_3|^2+m^2_Z \cos 2 \beta \left(\frac{1}{2}-\frac{2}{3}\sin^2\theta_W\right),  \nonumber\\
m^2_{88}&= M^2_{\tilde t_3}+\frac{v^2_2|y'_4|^2}{2}+|h''_5|^2   +\frac{2}{3}m^2_Z \cos 2 \beta \sin^2\theta_W\ . \nonumber
\end{align}

\begin{align}
m^2_{12}&=m^{2*}_{21}=-\frac{ v_1 y_2h^*_3}{\sqrt{2}} +\frac{ v_2 h_5 y'^*_1}{\sqrt{2}} ,
m^2_{13}=m^{2*}_{31}=\frac{y'^*_1}{\sqrt{2}}(v_2 A^*_{t} -\mu v_1),
m^2_{14}=m^{2*}_{41}=0,\nonumber\\
 m^2_{15} &=m^{2*}_{51}=h'_3 h^*_3,
 m^{2*}_{16}= m^{2*}_{61}=0,  m^{2*}_{17}= m^{2*}_{71}=h''_3 h^*_3,  m^{2*}_{18}= m^{2*}_{81}=0,\nonumber\\
m^2_{23}&=m^{2*}_{32}=0,
m^2_{24}=m^{2*}_{42}=\frac{y^*_2}{\sqrt{2}}(v_1 A^*_{T} -\mu v_2),  m^2_{25} = m^{2*}_{52}= -\frac{ v_1 h'_3y^*_2}{\sqrt{2}} +\frac{ v_2 y'_3 h'^*_5}{\sqrt{2}} ,\nonumber\\
 m^2_{26} &=m^{2*}_{62}=0,  m^2_{27} =m^{2*}_{72}=  -\frac{ v_1 h''_3y^*_2}{\sqrt{2}} +\frac{ v_2 y'_4 h''^*_5}{\sqrt{2}},  m^2_{28} =m^{2*}_{82}=0, \nonumber\\
m^2_{34}&=m^{2*}_{43}= \frac{ v_1 h_5 y^*_2}{\sqrt{2}} -\frac{ v_2 y'_1 h^*_3}{\sqrt{2}}, m^2_{35} =m^{2*}_{53} =0, m^2_{36} =m^{2*}_{63}=h_5 h'^*_5,\nonumber\\
 m^2_{37} &=m^{2*}_{73} =0,  m^2_{38} =m^{2*}_{83} =h_5 h''^*_5,\nonumber\\
m^2_{45}&=m^{2*}_{54}=0, m^2_{46}=m^{2*}_{64}=-\frac{y'^*_3 v_2 h'_3}{\sqrt{2}}+\frac{v_1 y_2 h'^*_5}{\sqrt{2}},
\nonumber\\
%\end{align}
m^2_{47}&=m^{2*}_{74}=0,
m^2_{48}=m^{2*}_{84}=\frac{v_1 y_2 h''^*_5}{\sqrt{2}}-\frac{v_2 y'^*_4 h''_3}{\sqrt{2}},\nonumber\\
 m^2_{56}&=m^{2*}_{65}=\frac{y'^*_3}{\sqrt{2}}(v_2 A^*_{c}-\mu v_1), \nonumber\\
m^2_{57}&=m^{2*}_{75}= h''_3 h'^*_3, m^2_{58}=m^{2*}_{85}=0, \nonumber\\
m^2_{67}&=m^{2*}_{76}=0, m^2_{68}=m^{2*}_{86}= h'_5 h''^*_5, \nonumber\\
m^2_{78}&=m^{2*}_{87}=\frac{y'^*_4}{\sqrt{2}}(v_2 A^*_{u}-\mu v_1).
%\label{15}
\end{align}

We can diagonalize the sneutrino mass square matrix  by the  unitary transformation
\begin{equation}
 \tilde D^{u\dagger} M^2_{\tilde u} \tilde D^{u} = \text{diag} (M^2_{\tilde u_1}, M^2_{\tilde u_2}, M^2_{\tilde u_3},  M^2_{\tilde u_4},M^2_{\tilde u_5},  M^2_{\tilde u_6}, M^2_{\tilde u_7}, M^2_{\tilde u_8})\ .
\end{equation}


\newpage

\begin{thebibliography}{999}

%\cite{Golub:1994cg}
\bibitem{Golub:1994cg}
  R.~Golub and K.~Lamoreaux,
  %``Neutron electric dipole moment, ultracold neutrons and polarized He-3,''
  Phys.\ Rept.\  {\bf 237}, 1 (1994).
  %%CITATION = PRPLC,237,1;%%
  %75 citations counted in INSPIRE as of 22 mar 2015

\bibitem{sm}
W.~Bernreuther and M. ~Suzuki, Rev. Mod. Phys. 63,
313 (1991);
I.I.Y. Bigi  and N. G. Uraltsev, Sov. Phys. JETP 73,
198 (1991);
M. J. Booth, eprint hep-ph/9301293;
Gavela, M. B., et al., Phys. Lett. B109, 215 (1982);
I. B. Khriplovich and A. R. Zhitnitsky,  Phys. Lett.
B109, 490 (1982);
E. P. Shabalin, Sov. Phys. Usp. 26, 297 (1983);
I.~B.~ Kriplovich and S.~K.~ Lamoureaux, {\it CP Violation Without Strangeness},
 (Springer, 1997).

\bibitem{Ibrahim:2007fb}
  T.~Ibrahim and P.~Nath,
 % ``CP violation from Standard Model to strings,''
  Rev.\ Mod.\ Phys.\  {\bf 80}, 577 (2008);
 % [arXiv:0705.2008 [hep-ph];
  %%CITATION = RMPHA,80,577;%%
%T.~Ibrahim and P.~Nath,
 % ``Phases and CP violation in SUSY,''
  arXiv:hep-ph/0210251.
  %%CITATION = HEP-PH/0210251;%%
    A.~Pilaftsis,
  %``CP violation in the Higgs sector of the MSSM,''
  hep-ph/9908373;
  %%CITATION = HEP-PH/9908373;%%
 M.~Pospelov and A.~Ritz,
  %``Electric dipole moments as probes of new physics,''
  Annals Phys.\  {\bf 318}, 119 (2005)
  [hep-ph/0504231];
  %%CITATION = HEP-PH/0504231;%%
  J.~Engel, M.~J.~Ramsey-Musolf and U.~van Kolck,
  %``Electric Dipole Moments of Nucleons, Nuclei, and Atoms: The Standard Model and Beyond,''
  Prog.\ Part.\ Nucl.\ Phys.\  {\bf 71}, 21 (2013)
  [arXiv:1303.2371 [nucl-th]].
  %%CITATION = ARXIV:1303.2371;%%


%\cite{Hewett:2012ns}
\bibitem{Hewett:2012ns}
  J.~L.~Hewett, H.~Weerts, R.~Brock, J.~N.~Butler, B.~C.~K.~Casey, J.~Collar, A.~de Gouvea and R.~Essig {\it et al.},
  %``Fundamental Physics at the Intensity Frontier,''
  arXiv:1205.2671 [hep-ex].
  %%CITATION = ARXIV:1205.2671;%%
  %155 citations counted in INSPIRE as of 22 mar 2015


\bibitem{earlywork} J. ~Ellis, S. ~Ferrara, and D.~V. ~Nanopoulos,
Phys. Lett.  {\bf 114B} (1982) 231;
 W.~Buchmuller and D.~Wyler, Phys. Lett. B121 (1983) 321;
  F.~del'Aguila, M.~B. ~Gavela, J.~A.~Grifols and A.~Mendez, Phys. Lett. B126 (1983) 71;
 J.~Polchinski and M.~B.~Wise, Phys.~Lett. B125 (1983) 393;
 E.~Franco and M.~Mangano, Phys. Lett. B135 (1984) 445.

%\cite{Nath:1991dn}
\bibitem{Nath:1991dn}
  P.~Nath,
  %``CP Violation via electroweak gauginos and the electric dipole moment of the electron,''
  Phys.\ Rev.\ Lett.\  {\bf 66}, 2565 (1991);
  %%CITATION = PRLTA,66,2565;%%
  %189 citations counted in INSPIRE as of 22 mar 2015

\bibitem{cancellation1}
 T.~Ibrahim and P.~Nath,
  %``The Chromoelectric and purely gluonic operator contributions to the neutron electric dipole moment in N=1 supergravity,''
  Phys.\ Lett.\ B {\bf 418}, 98 (1998)
  [hep-ph/9707409];
  %%CITATION = HEP-PH/9707409;%%
  %``The Neutron and the electron electric dipole moment in N=1 supergravity unification,''
  Phys.\ Rev.\ D {\bf 57}, 478 (1998)
   [hep-ph/9708456];
  %%CITATION = HEP-PH/9708456;%%
   %``The Neutron and the lepton EDMs in MSSM, large CP violating phases, and the cancellation mechanism,''
  Phys.\ Rev.\ D {\bf 58}, 111301 (1998)
  [hep-ph/9807501];
  %%CITATION = HEP-PH/9807501;%%
   %``Large CP phases and the cancellation mechanism in EDMs in SUSY, string and brane models,''
  Phys.\ Rev.\ D {\bf 61}, 093004 (2000)
  [hep-ph/9910553].
  %%CITATION = HEP-PH/9910553;%%

  \bibitem{cancellation2}
     T.~Falk and K.~A.~Olive,
  %``More on electric dipole moment constraints on phases in the constrained MSSM,''
  Phys.\ Lett.\ B {\bf 439}, 71 (1998)
  [hep-ph/9806236];
  %%CITATION = HEP-PH/9806236;%%
   M.~Brhlik, G.~J.~Good and G.~L.~Kane,
  %``Electric dipole moments do not require the CP violating phases of supersymmetry to be small,''
  Phys.\ Rev.\ D {\bf 59}, 115004 (1999)
  [hep-ph/9810457].
  %%CITATION = HEP-PH/9810457;%%
  %347 citations counted in INSPIRE as of 11 Apr 2014

%\cite{Babu:1999xf}
\bibitem{Babu:1999xf}
  K.~S.~Babu, B.~Dutta and R.~N.~Mohapatra,
  %``Seesaw constrained MSSM, solution to the SUSY CP problem and a supersymmetric explanation of epsilon-prime / epsilon,''
  Phys.\ Rev.\ D {\bf 61}, 091701 (2000)
  [hep-ph/9905464].
  %%CITATION = HEP-PH/9905464;%%
  %100 citations counted in INSPIRE as of 22 mar 2015


%\cite{McKeen:2013dma}
\bibitem{McKeen:2013dma}
  D.~McKeen, M.~Pospelov and A.~Ritz,
  %``Electric dipole moment signatures of PeV-scale superpartners,''
  Phys.\ Rev.\ D {\bf 87}, no. 11, 113002 (2013)
  [arXiv:1303.1172 [hep-ph]].
  %%CITATION = ARXIV:1303.1172;%%
  %34 citations counted in INSPIRE as of 22 mar 2015


%\cite{Moroi:2013sfa}
\bibitem{Moroi:2013sfa}
  T.~Moroi and M.~Nagai,
  %``Probing Supersymmetric Model with Heavy Sfermions Using Leptonic Flavor and CP Violations,''
  Phys.\ Lett.\ B {\bf 723}, 107 (2013)
  [arXiv:1303.0668 [hep-ph]].
  %%CITATION = ARXIV:1303.0668;%%
  %28 citations counted in INSPIRE as of 22 mar 2015


%\cite{Altmannshofer:2013lfa}
\bibitem{Altmannshofer:2013lfa}
  W.~Altmannshofer, R.~Harnik and J.~Zupan,
  %``Low Energy Probes of PeV Scale Sfermions,''
  JHEP {\bf 1311}, 202 (2013)
  [arXiv:1308.3653 [hep-ph]].
  %%CITATION = ARXIV:1308.3653;%%
  %27 citations counted in INSPIRE as of 22 mar 2015


%\cite{Ibrahim:2014tba}
\bibitem{Ibrahim:2014tba}
  T.~Ibrahim, A.~Itani and P.~Nath,
  %``Electron electric dipole moment as a sensitive probe of PeV scale physics,''
  Phys.\ Rev.\ D {\bf 90}, no. 5, 055006 (2014).
  %%CITATION = PHRVA,D90,055006;%%
  %3 citations counted in INSPIRE as of 22 mar 2015


%\cite{Dhuria:2013ida}
\bibitem{Dhuria:2013ida}
  M.~Dhuria and A.~Misra,
  %``A Healthy Electron/Neutron EDM in D3/D7 mu-Split SUSY,''
  arXiv:1308.3233 [hep-ph].
  %%CITATION = ARXIV:1308.3233;%%


%\cite{Baker:2006ts}
\bibitem{Baker:2006ts}
  C.~A.~Baker, D.~D.~Doyle, P.~Geltenbort, K.~Green, M.~G.~D.~van der Grinten, P.~G.~Harris, P.~Iaydjiev and S.~N.~Ivanov {\it et al.},
  %``An Improved experimental limit on the electric dipole moment of the neutron,''
  Phys.\ Rev.\ Lett.\  {\bf 97}, 131801 (2006)
  [hep-ex/0602020].
  %%CITATION = HEP-EX/0602020;%%
  %620 citations counted in INSPIRE as of 22 mar 2015


%\cite{Ito:2007xd}
\bibitem{Ito:2007xd}
  T.~M.~Ito,
  %``Plans for a Neutron EDM Experiment at SNS,''
  J.\ Phys.\ Conf.\ Ser.\  {\bf 69}, 012037 (2007)
  [nucl-ex/0702024 [NUCL-EX]].
  %%CITATION = NUCL-EX/0702024;%%
  %36 citations counted in INSPIRE as of 22 mar 2015


\bibitem{vectorlike}
 H.~Georgi,
  %``Towards A Grand Unified Theory Of Flavor,''
  Nucl.\ Phys.\  B {\bf 156}, 126 (1979);
  %%CITATION = NUPHA,B156,126;%%
  F.~Wilczek and A.~Zee,
  %``Families From Spinors,''
  Phys.\ Rev.\  D {\bf 25}, 553 (1982);
  %%CITATION = PHRVA,D25,553;%%
J. Maalampi, J.T. Peltoniemi, and M. Roos, PLB 220, 441(1989);
  J.~Maalampi and M.~Roos,
 % ``Physics Of Mirror Fermions,''
  Phys.\ Rept.\  {\bf 186}, 53 (1990);
  %%CITATION = PRPLC,186,53;%%
  K.~S.~Babu, I.~Gogoladze, P.~Nath and R.~M.~Syed,
  % ``A Unified framework for symmetry breaking in SO(10),''
  Phys.\ Rev.\ D {\bf 72}, 095011 (2005)
  [hep-ph/0506312];
  %%CITATION = HEP-PH/0506312;%%
 % ``Fermion mass generation in SO(10) with a unified Higgs sector,''
  Phys.\ Rev.\  D {\bf 74}, 075004 (2006),
  [arXiv:hep-ph/0607244];
  %%CITATION = PHRVA,D74,075004;%%
 % ``Variety of SO(10) GUTs with Natural Doublet-Triplet Splitting via the Missing Partner Mechanism,''
  Phys.\ Rev.\ D {\bf 85}, 075002 (2012)
  [arXiv:1112.5387 [hep-ph]];
  %%CITATION = ARXIV:1112.5387;%%
   P.~Nath and R.~M.~Syed,
  %``Yukawa Couplings and Quark and Lepton Masses in an SO(10) Model with a
 % Unified Higgs Sector,''
  Phys.\ Rev.\  D {\bf 81}, 037701 (2010).
%  [arXiv:0909.2380 [hep-ph]].
  %%CITATION = PHRVA,D81,037701;%%


%\cite{Babu:2008ge}
\bibitem{Babu:2008ge}
  K.~S.~Babu, I.~Gogoladze, M.~U.~Rehman and Q.~Shafi,
  %``Higgs Boson Mass, Sparticle Spectrum and Little Hierarchy Problem in Extended MSSM,''
  Phys.\ Rev.\ D {\bf 78}, 055017 (2008)
  [arXiv:0807.3055 [hep-ph]].
  %%CITATION = ARXIV:0807.3055;%%
  %91 citations counted in INSPIRE as of 22 mar 2015


%\cite{Liu:2009cc}
\bibitem{Liu:2009cc}
  C.~Liu,
  %``Supersymmetry and Vector-like Extra Generation,''
  Phys.\ Rev.\ D {\bf 80}, 035004 (2009)
  [arXiv:0907.3011 [hep-ph]].
  %%CITATION = ARXIV:0907.3011;%%
  %28 citations counted in INSPIRE as of 22 mar 2015


%\cite{Martin:2009bg}
\bibitem{Martin:2009bg}
  S.~P.~Martin,
  %``Extra vector-like matter and the lightest Higgs scalar boson mass in low-energy supersymmetry,''
  Phys.\ Rev.\ D {\bf 81}, 035004 (2010)
  [arXiv:0910.2732 [hep-ph]].
  %%CITATION = ARXIV:0910.2732;%%
  %130 citations counted in INSPIRE as of 22 mar 2015


%\cite{Ibrahim:2011im}
\bibitem{Ibrahim:2011im}
  T.~Ibrahim and P.~Nath,
  %``The Chromoelectric Dipole Moment of the Top Quark in Models with Vector Like Multiplets,''
  Phys.\ Rev.\ D {\bf 84}, 015003 (2011)
  [arXiv:1104.3851 [hep-ph]].
  %%CITATION = ARXIV:1104.3851;%%
  %15 citations counted in INSPIRE as of 22 mar 2015


%\cite{Ibrahim:2010hv}
\bibitem{Ibrahim:2010hv}
  T.~Ibrahim and P.~Nath,
  %``The Top quark electric dipole moment in an MSSM extension with vector like multiplets,''
  Phys.\ Rev.\ D {\bf 82}, 055001 (2010)
  [arXiv:1007.0432 [hep-ph]].
  %%CITATION = ARXIV:1007.0432;%%
  %18 citations counted in INSPIRE as of 22 mar 2015


%\cite{Ibrahim:2010va}
\bibitem{Ibrahim:2010va}
  T.~Ibrahim and P.~Nath,
  %``Large Tau and Tau Neutrino Electric Dipole Moments in Models with Vector Like Multiplets,''
  Phys.\ Rev.\ D {\bf 81}, no. 3, 033007 (2010)
  [Erratum-ibid.\ D {\bf 89}, no. 11, 119902 (2014)]
  [arXiv:1001.0231 [hep-ph]].
  %%CITATION = ARXIV:1001.0231;%%
  %19 citations counted in INSPIRE as of 22 mar 2015


%\cite{Ibrahim:2008gg}
\bibitem{Ibrahim:2008gg}
  T.~Ibrahim and P.~Nath,
  %``An MSSM Extension with a Mirror Fourth Generation, Neutrino Magnetic Moments and LHC Signatures,''
  Phys.\ Rev.\ D {\bf 78}, 075013 (2008)
  [arXiv:0806.3880 [hep-ph]].
  %%CITATION = ARXIV:0806.3880;%%
  %22 citations counted in INSPIRE as of 22 mar 2015


%\cite{Ibrahim:2009uv}
\bibitem{Ibrahim:2009uv}
  T.~Ibrahim and P.~Nath,
  %``On the Possible Observation of Mirror Matter,''
  Nucl.\ Phys.\ Proc.\ Suppl.\  {\bf 200-202}, 161 (2010)
  [arXiv:0910.1303 [hep-ph]].
  %%CITATION = ARXIV:0910.1303;%%
  %11 citations counted in INSPIRE as of 22 mar 2015


%\cite{Ibrahim:2012ds}
\bibitem{Ibrahim:2012ds}
  T.~Ibrahim and P.~Nath,
  %``$\tau\to \mu \gamma$ decay in extensions with a vectorlike generation,''
  Phys.\ Rev.\ D {\bf 87}, no. 1, 015030 (2013)
  [arXiv:1211.0622 [hep-ph]].
  %%CITATION = ARXIV:1211.0622;%%
  %7 citations counted in INSPIRE as of 22 mar 2015


%\cite{Ibrahim:2014oia}
\bibitem{Ibrahim:2014oia}
  T.~Ibrahim, A.~Itani and P.~Nath,
  %``Electron EDM as a Sensitive Probe of PeV Scale Physics,''
  arXiv:1406.0083 [hep-ph].
  %%CITATION = ARXIV:1406.0083;%%
  %2 citations counted in INSPIRE as of 22 mar 2015


%\cite{Aboubrahim:2014hya}
\bibitem{Aboubrahim:2014hya}
  A.~Aboubrahim, T.~Ibrahim and P.~Nath,
  %``Probe of New Physics using Precision Measurement of the Electron Magnetic Moment,''
  Phys.\ Rev.\ D {\bf 89}, no. 9, 093016 (2014)
  [arXiv:1403.6448 [hep-ph]].
  %%CITATION = ARXIV:1403.6448;%%
  %4 citations counted in INSPIRE as of 22 mar 2015


%\cite{Aboubrahim:2013yfa}
\bibitem{Aboubrahim:2013yfa}
  A.~Aboubrahim, T.~Ibrahim, A.~Itani and P.~Nath,
  %``Large Neutrino Magnetic Dipole Moments in MSSM Extensions,''
  Phys.\ Rev.\ D {\bf 89}, no. 5, 055009 (2014)
  [arXiv:1312.2505 [hep-ph]].
  %%CITATION = ARXIV:1312.2505;%%
  %7 citations counted in INSPIRE as of 22 mar 2015


%\cite{Aboubrahim:2013gfa}
\bibitem{Aboubrahim:2013gfa}
  A.~Aboubrahim, T.~Ibrahim and P.~Nath,
  %``Radiative Decays of Cosmic Background Neutrinos in Extensions of MSSM with a Vector Like Lepton Generation,''
  Phys.\ Rev.\ D {\bf 88}, 013019 (2013)
  [arXiv:1306.2275 [hep-ph]].
  %%CITATION = ARXIV:1306.2275;%%
  %7 citations counted in INSPIRE as of 22 mar 2015

%\cite{Tsentalovich:2014mfa}
\bibitem{Tsentalovich:2014mfa}
  E.~P.~Tsentalovich [nEDM Collaboration],
  %``The nEDM experiment at the SNS,''
  Phys.\ Part.\ Nucl.\  {\bf 45}, 249 (2014).
  %%CITATION = PPNUE,45,249;%%

%\cite{Y.V.Stadnik:2014}
\bibitem{Y.V.Stadnik:2014}
Y. V. Stadnik and V. V. Flambaum,
Phys.\ Rev.\ D {\bf 89}, 043522 (2014). 
  
\end{thebibliography}
\end{document}